\documentclass[11pt, a4paper]{article}
\usepackage{authblk}
\usepackage[greek,english]{babel}

\usepackage[T1]{fontenc}
\usepackage[utf8]{inputenc}
\usepackage{kerkis}
\usepackage{alphabeta}

\usepackage{listings}
\usepackage{fancyvrb}
\usepackage{listings}
\usepackage{tabularx} 
\usepackage{amsmath,bm}  
\usepackage{graphicx} 
\usepackage[a4paper,left=2.5cm,right=2.5cm,top=2.5cm,bottom=2.5cm]{geometry} 
\usepackage{cite} 
\usepackage[final]{hyperref} 
\hypersetup{
	colorlinks=true,       
	linkcolor=blue,        
	citecolor=[rgb]{0,0.1,0.5} ,       
	urlcolor=green         
}
\usepackage{subfig}
\graphicspath{ {images/} }
\usepackage{soul,color}

\addto\captionsenglish{%
  }

\title{\includegraphics[scale = 0.3]{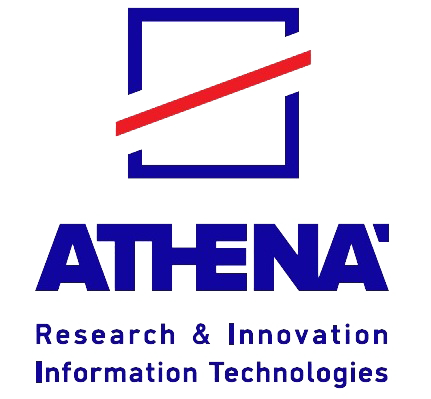} 
       \\ \textbf{Οπτικοακουστική Σύνθεση Φωνής}
       \\ \Large{Σύντομη Βιβλιογραφική Ανασκόπηση}
       }
\author{Ευθύμιος Γεωργίου και Αθανάσιος Κατσαμάνης}
\affil{Ινστιτούτο Επεξεργασίας του Λόγου, Ε.Κ. ΑΘΗΝΑ}
\date{}

\begin{document}
\maketitle

\begin{abstract}
    Η παρούσα βιβλιογραφική ανασκόπηση μελετάει το πρόβλημα της οπτικοακουστικής σύνθεσης φωνής. Ουσιαστικά δηλαδή εξετάζουμε πως μπορούμε από κάποιο κείμενο εισόδου να συνθέσουμε μια ανθρωπόμορφη οπτική ροή καθώς επίσης και την αντίστοιχη φωνή. Εξαιτίας της πολύ μεγάλης πολυπλοκότητας του προβλήματος αυτού, χρειάζεται να το μελετήσουμε σε δύο επιμέρους τμήματα. Συγκεκριμένα, αυτό της σύνθεσης φωνής από κείμενο (text-to-speech synthesis) καθώς και τη σύνθεση ανθρωπόμορφης ροής από φωνή. Σε ότι αφορά τη σύνθεση φωνής μελετάμε τόσο τα δίκτυα που κάνουν την απεικόνιση από το κείμενο σε κάποια ενδιάμεση αναπράσταση καθώς επίσης και τα δίκτυα που παράγουν φωνή από τις ενδιάμεσες αυτές αναπραστάσεις. Ως προς την οπτική σύνθεση, κατηγοριοποιούμε τις προσεγγίσεις με βάση το αν παράγουν ανθρώπινα πρόσωπα ή ανθρωπόμορφες φιγούρες. Προσπάθεια γίνεται επίσης να παρουσιαστεί η σημασία της επιλογής των μοντέλων προσώπου στη δεύτερη περίπτωση. Καθόλη την έκταση της ανασκόπησης, παρουσιάζουμε τις σημαντικότερες, κατά τη γνώμη μας, εργασίες και στα δύο αυτά πεδία, προσπαθώντας να δώσουμε βάση στα πλεονεκτήματα και μειονεκτήματα της κάθε μιας.
\end{abstract}


\newpage

\tableofcontents

\newpage

\section{Εισαγωγή}
H αλληλεπίδραση ανθρώπου-μηχανής (human computer interaction) στον $21^{ο}$ αιώνα αυξάνεται με ραγδαίους ρυθμούς, εξαιτίας της συνεχόμενης τεχνολογικής εξέλιξης. Ειδικότερα σε καθημερινό πλέον επίπεδο , σε μικρότερο ή μεγαλύτερο βαθμό, έχουμε επαφή με εικονικούς, φυσικούς ή έξυπνους πράκτορες (agents\footnote{Ως πράκτορας (agent) νοείται μια οντότητα η οποία αντιλαμβάνεται το περιβάλλον της μέσω αλληλεπίδρασης με αυτό. Παράδειγμα φυσικού agent είναι ένα ρομπότ ενώ εικονικού ένα avatar.}). Πιο συγκεκριμένα αυτοί οι agents συμμετέχουν σε πληθώρα εφαρμογών από καθημερινές δραστηριότητες (chatbots, πλοήγηση στο διαδίκτυο) μέχρι και παιδαγωγικούς σκοπούς (learning assistants) \cite{johnson2000animated}. 
\par
Η τεχνητή νοημοσύνη (artificial intelligence) αποσκοπεί στο να μεγιστοποιήσει τη φυσικότητα στην αλληλεπίδραση ανθρώπου μηχανής. Λαμβάνοντας υπόψιν το γεγονός ότι η βασική συνιστώσα επικοινωνίας είναι η ομιλία και συνεπακόλουθα η ανθρώπινη φωνή, αντιλαμβανόμαστε το πόσο ζωτικής σημασίας είναι η ανάπτυξη αλγορίθμων που παράγουν ρεαλιστική ανθρώπινη φωνή. Λόγω όμως της πολυτροπικής (multimodal) φύσης της φωνής \cite{ekman1984expression} εξίσου σημαντικό είναι να παράξουμε και οπτική φωτορεαλιστική πληροφορία. Έχει αποδειχθεί άλλωστε από μελετητές πως ο συνδυασμός οπτικής και φωνητικής πληροφορίας επιφέρει τα καλύτερα δυνατά αποτελέσματα ως προς την ευρωστία της επικοινωνίας \cite{sumby1954visual}.
\par
Η παρούσα βιβλιογραφική ανασκόπηση πραγματεύεται το πρόβλημα της \textit{οπτικοακουστικής σύνθεσης φωνής} (audio-visual speech synthesis). Σκοπός του υπό εξέταση προβλήματος είναι δοθείσας γραπτής περιγραφής να παράξουμε ένα φωτο-ρεαλιστικό βίντεο το οποίο θα συνοδεύεται από ένα επιπλέον σήμα φωνής. Φυσικά τόσο η οπτική όσο και η ακουστική συνιστώσα που συντίθενται θα πρέπει να είναι συγχρονισμένες.

\subsection{Περιγραφή του Προβλήματος}
Μετατοπίζοντας την παρουσίαση σε πιο τεχνικό επίπεδο μπορούμε να κάνουμε τις ακόλουθες γενικές παρατηρήσεις. Όπως ίσως είναι φανερό το πρόβλημα της οπτικοακουστικής σύνθεσης φωνής διαιρείται σε δύο υπό-προβλήματα και συγκεκριμένα τη σύνθεση
\begin{itemize}
    \item φωνής από κείμενο (text-to-speech synthesis)
    \item οπτικής συνιστώσας από ηχητικό σήμα
\end{itemize}
Και τα δύο προβλήματα έχουν μελετηθεί στη βιβλιογραφία αλλά το ενδιαφέρον των ερευνητών είναι σαφώς προσανατολισμένο στη σύνθεση φωνής
(speech synthesis) σε σχέση με τα δύο. 
\par 
Είναι γεγονός πως από τη βιβλιογραφία απουσιάζουν εργασίες οι οποίες μελετάνε καθολικά το πρόβλημα που θέλουμε να επιλύσουμε. Αυτό οφείλεται εν πολλοίς στην πολυπλοκότητα του συνδυαστικού προβλήματος που θέλουμε να επιλύσουμε.
Συγκεκριμένα, ακόμα και οι δύο γενικές κατευθύνσεις που σκιαγραφήσαμε προηγουμένως, χωρίζονται σε επιμέρους περιοχές κάθε μία από τις οποίες χρήζει ιδιαίτερης ερευνητικής μεταχείρισης. Στη συνέχεια παρουσιάζουμε μια βιβλιογραφική ανασκόπηση των μεθόδων και αλγορίθμων επί των κατευθύνσεων αυτών.

\subsection{Σύνθεση Φωνής}
Με τη ραγδαία εξέλιξη της βαθιάς μηχανικής μάθησης (deep learning) ως ερευνητικού πεδίου πολλά προβλήματα επιλύθηκαν πολύ αποδοτικότερα απ'ότι με παλαιότερες προσεγγίσεις \cite{krizhevsky_imagenet_2012, lecun_gradient-based_1998}. 
\par
\begin{figure}[!htb]
    \centering
    \includegraphics[scale=0.4]{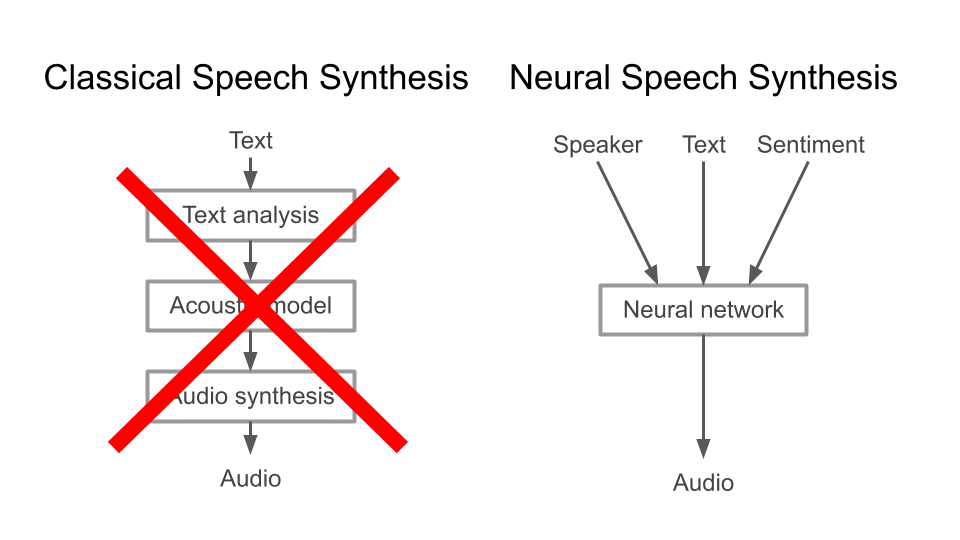}
    \caption{Κλασική και Νευρωνική προσέγγιση Σύνθεσης Φωνής}
    \label{fig:tts}
\end{figure}

Ένα από αυτά τα προβλήματα στα οποία η τεχνητή νοημοσύνη ξεπέρασε κατά πολύ τις προηγούμενες προσεγγίσεις και πλησίσασε τα όρια της \href{https://deepmind.com/blog/article/wavenet-generative-model-raw-audio}{ανθρώπινης διαίσθησης} είναι και η σύνθεση φωνής \cite{oord_wavenet_2016}. 
Oι παλαιότερες προσεγγίσεις βασίζονται στην ανάλυση κειμένου με σκοπό την εξαγωγή κάποιων γλωσσικών χαρακτηριστικών τα οποία δίνονται ως είσοδο σε κάποιο ακουστικό μοντέλο (acoustic model), π.χ GMMs. Το acoustic model με τη σειρά του τροφοδοτεί την έξοδό του, συνήθως φασματογραφήματα (spectrograms), σε ένα σύστημα σύνθεσης φωνής το οποίο παράγει την τελική συνθετική φωνή.
Η προαναφερθείσα διαδικασία φαίνεται και στο Σχήμα \ref{fig:tts}.
\par
Μεταβαίνοντας τώρα σε προσεγγίσεις με βάση τα νευρωνικά δίκτυα, η βασική ιδέα είναι να αντικαταστήσουμε τα επιμέρους υποσυστήματα είτε με μια ενιαία είτε με πολλαπλές αρχιτεκτονικές.
Για παράδειγμα είναι σύνηθες \cite{wang_tacotron_2017} ένα υποδίκτυο της συνολικής αρχιτεκτονικής να αναλαμβάνει να απεικονίσει τη λεκτική περιγραφή (κείμενο) σε καποιου είδους spectrogram και στη συνέχεια να χρησιμοποιηθεί ένας \textit{vocoder}, ο οποίος αναλαμβάνει να ανακατσκευάσει το ηχητικό σήμα από τη φασματική του αναπαράσταση.

\subsection{Oπτικοακουστική Σύνθεση Φωνής}
Όταν αναφερόμαστε στο πρόβλημα της οπτικοακουστικής σύνθεσης φωνής τότε ιδανικά αυτό που θέλουμε να πετύχουμε είναι να κατασκευάσουμε έναν αλγόριθμο ο οποίος θα μπορεί να εκφέρεται με τον ίδιο τρόπο που θα το έκανε και ένας άνθρωπος. Αυτό σημαίνει ότι θα πρέπει αφενός να μπορεί να εκδηλώνει το τι θέλει να πει αφετέρου να κάνει τις κατάλληλες κινήσεις που χρησιμοποιούνε και οι ίδιοι οι άνθρωποι.
\par
Εξετάζοντας το πρόβλημα λίγο πιο προσεκτικά, μπορούμε να πούμε πως
αρχικά, η (παραγόμενη) φωνή σε συνδυασμό με τις εκφράσεις του προσώπου παρότι είναι αρκετά συσχετισμένες (correlated) σαν έννοιες, βρίσκονται στην πραγματικότητα σε δύο τελείως διαφορετικούς χώρους.
Συνεπώς θα πρέπει να βρούμε κάποιον (μη-γραμμικό) μετασχηματισμό που να μας απεικονίζει από τον ακουστικό στον οπτικό χώρο ανάλογα με το τι θέλουμε να πούμε.
Συνολικά θα πρέπει να έχουμε ένα δίκτυο το οποίο θα αναλαμβάνει να μαθαίνει τέτοιους μετασχηματισμούς.
\par 
\begin{figure}[!htb]
    \centering
    \includegraphics[scale=0.45]{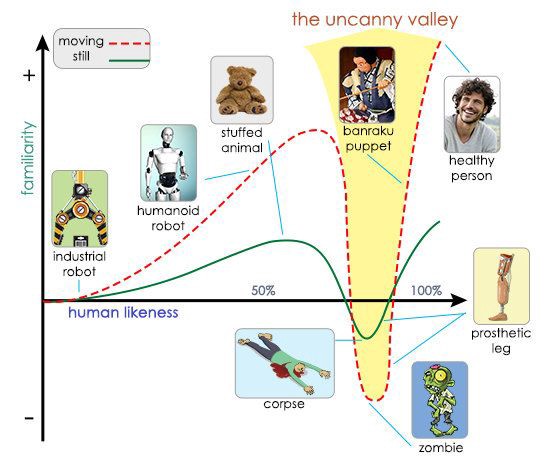}
    \caption{Uncanny Valley Effect (\href{https://areewitoelar.medium.com/the-uncanny-valley-of-improv-446befa2b89c}{Source})}
    \label{fig:uncanny}
\end{figure}
Το ακουστικό σήμα περιγράφεται είτε απυεθείας ως κυματομορφή είτε μέσω φασματικών αναπαραστάσεων.
Ως προς την οπτική συνιστώσα τώρα θα χρειαστεί να κατασκευάσουμε μια ανθρωπόμορφη κεφαλή (cartoon, avatar) η οποία να κινείται σε συμφωνία με το ηχητικό σήμα.
Ιδιαίτερη έμφαση πρέπει να δωθεί σε αυτό το πρόβλημα καθώς εξαιτίας της υψηλής μεταβλητότητας και πολυπλοκότητας που παρουσιάζεται, υπάρχει η πιθανότητα να προκληθεί το λεγόμενο uncanny valley effect\footnote{Εξαιτίας του εξελιγμένου οπτικού συστήματος που έχει αναπτύξει ο άνθρωπος ατέλειες αναφορικά με την οπτική συνιστώσα γίνονται πολύ έυκολα αντιληπτές.} \cite{seyama2007uncanny, mori2012uncanny}.

\section{Το πρόβλημα μετατροπής κειμένου σε φωνή}
Το πρόβλημα της σύνθεσης φωνής από κείμενο ή αλλιώς Text-to-Speech (TTS) Synthesis είναι ένα αρκετά δύσκολο πρόβλημα που καλείται να επιλύσει ταυτόχρονα τρία σημαντικά προβλήματα. Αρχικά μια οπτική του είναι ότι αποτελεί ένα πρόβλημα \textit{αποσυμπίεσης} (decompression) αφού ουσιαστικά θέλει να απεικονίσει περιορισμένους γραφικούς-κειμενικούς χαρακτήρες σε μια κυματομορφή η οποία αποτελείται από αρκετές χιλιάδες δείγματα. Επιπλέον, θα πρέπει να αντιστοιχίσουμε κάθε μονάδα κειμένου σε ένα τμήμα μια κυματομορφής συνεπώς λύνοντας ταυτόχρονα και ένα πρόβλημα \textit{συγχρονισμού-ευθυγράμμισης} (alignment) των δύο συνιστωσών. Τρίτον, καλούμαστε να επιλύσουμε το ευρύτετο πρόβλημα \textit{απεικόνισης μιας ακολουθίας σε μία άλλη}, το οποίο στη βιβλιογραφία αναφέρεται ως translation πρόβλημα.
\par
Στην ενότητα αυτή θα επικτρωθούμε στην περιγραφή προσεγγίσεων του προβλήματος σύνθεσης φωνής με χρήση βαθιών νευρωνικών δικτύων.
Συγκεκριμένα οι προσεγγίσεις αυτές μπορούν να χωριστούν σε δύο επιμέρους κατηγορίες, ανάλογα με το μέρος της διαδικασίας της σύνθεσης που εστιάζουν. Συγκεκριμένα χωρίζονται:
\begin{itemize}
    \item σε αυτές που λύνουν το πρόβλημα της παραγωγής κυματομορφής από ακουστικά χαρακτηριστικά, δηλαδή vocoders
    \item σε αυτές που απεικονίζουν τις λεξικές αναπαραστάσεις σε φασματικό περιεχόμενο και συνήθως συνοδεύονται από κάποιο vocoder
\end{itemize}
Δεν είναι δύσκολο κανείς να δει την αναλογία μεταξύ των παραδοσιακών αλγορίθμων και των πιο σύγχρονων προσεγγίσεων που αφορούν την αναπαράσταση της κυματομορφής. 
Σε ότι αφορά την κατασκευή μιας end-to-end προσέγγισης, συνήθως χρησιμοποιούνται δύο δίκτυα.
Ένα που απεικονίζει το κείμενο σε ακουστική πληροφορία και έναν vocoder που κατασευάζει μια κυματομορφή βασιζόμενος σε αυτές.
Για το λόγο αυτό οι δύο αυτές προσεγγίσεις δεν θα πρέπει να θεωρούνται διαφορετικές, ή πόσο μάλλον αντιπαραθετικές, αλλά συμπληρωματικές καθώς η μία κάνει άμεση χρήση της άλλης.
Ο λόγος που το συνολικό πρόβλημα αντιμετωπίζεται σαν δύο επιμέρους υποπροβλήματα έγκειται στη δυσκολία και την πολυπλοκότητα του καθενός. Στη συνέχεια περιγράφουμε αναλυτικά τις δυο αυτές κατηγορίες.

\subsection{Neural Vocoders}
\begin{figure}[!htb]
    \centering
    \includegraphics[scale=0.25]{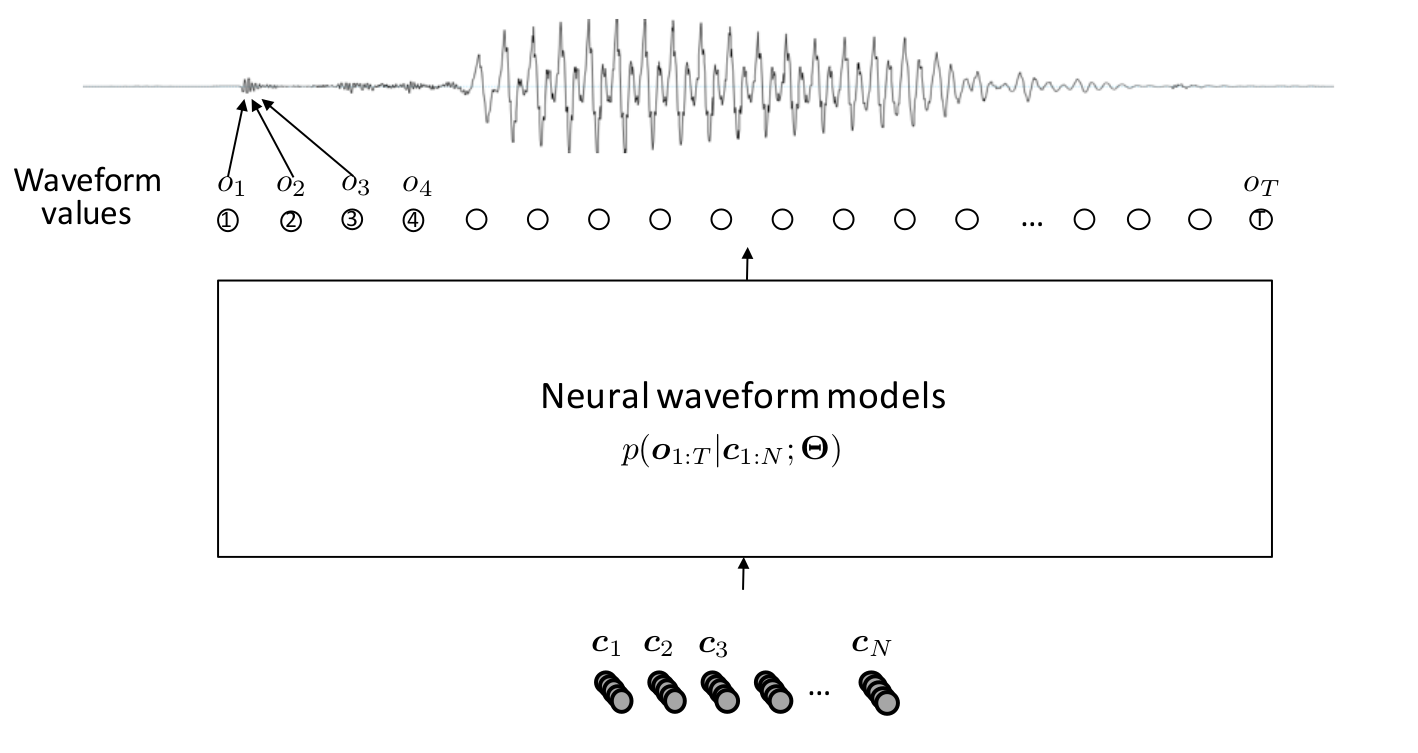}
    \caption{Neural Vocoders (\href{https://www.slideshare.net/jyamagis/tutorial-on-endtoend-texttospeech-synthesis-part-1-neural-waveform-modeling}{Source})}
    \label{fig:vocoder}
\end{figure}

Στην ενότητα αυτή παρουσιάζουμαι τις πιο ενδιαφέρουσες και συγχρόνως με υψηλές επιδόσεις προσεγγίσεις σχετικά με αρχιτεκτονικές που χρησιμοποιούνται ως neural vocoders. 
Όπως απεικονίζεται και στο Σχήμα \ref{fig:vocoder} τα δίκτυα αυτά αποσκοπούν στο να απεικονίσουν ακουστικά χαρακτηριστικά (acoustic features) $c_i$ στα δείγματα (τιμές) της αντίστοιχης κυματομορφής $o_i$. 
Η παράμετρος $\Theta$ συμβολίζει τις ``ελεύθερες" παραμέτρους του δικτύου τις οποίες καλούμαστε να εκπαιδεύσουμε μέσω των αλγορίθμων μάθησης \cite{rumelhart_learning_1986}.

Παρά το γεγονός ότι υπαχουν πολυάριθμες προσεγγίσεις, τα βασικά δομικά στοιχεία που τροποποιούνται είναι:
\begin{itemize}
    \item η αρχιτεκτονική του μοντέλου αυτού καθ'αυτού (π.χ dilated convolutions, normalizing flow, GAN)
    \item η/οι συνάρτηση/εις κόστους που χρησιμοποιούνται (loss functions)
    \item επιπλέον μετασχηματισμοί που μετασχηματίζουν το πρόβλημα σε κάποιο πιο απλό (knowledge-based transforms)
\end{itemize}

Αφού περιγράψαμε τα δομικά στοιχεία των προσεγγίσεων παραθέτουμε επιγραμματικά και τις επιμέρους προσεγγίσεις τις οποίες στη συνέχεια παρουσιάζουμε σε μεγαλύτερο βάθος. Αρχικά θα παρουσιάσουμε τις λεγόμενες αυτοπαλινδρομικές μεθόδους (\textit{autoregressive}) \cite{van2016conditional}, οι οποίες κάνουν χρήση dilated convolutions και είναι οι πρώτες που παρουσίασαν αποτελέσματα που προσεγγίζουν το ανθρώπινο σύστημα παραγωγής φωνής, από άποψη ποιότητας. Ακολουθούν οι μέθοδοι βασισμένες σε μοντέλα κανονικοποιημένης ροής (\textit{normalizing flow}) \cite{rezende2015variational, kingma2016improved, papamakarios2017masked}, οι οποίες μετασχηματίζουν το τελικό πρόβλημα σε πιο απλά εφαρμόζοντας διαδοχικούς μετασχηματισμούς σε μια κατανομή πιθανότητας και τέλος θα παρουσιάσουμε κάποιες μεθόδους που βασίζονται στα περίφημα Generative Adversarial Networks (GANs) \cite{goodfellow_generative_2014}.

\subsubsection{Autoregressive Models}
Η πρώτη δουλειά στο πεδίο αυτό, που εισήγαγε η DeepMind, είναι το περίφημο \textit{WaveNet} \cite{oord_wavenet_2016} το οποίο ουσιαστικά είναι ένα πλήρως πιθανοκρατικό (probabilistic) και autoregressive μοντέλο. Με άλλα λόγια, κάθε έξοδος είναι μια κατανομή πιθανότητας πάνω στα δυνατά δείγματα της κυματομορφής και ταυτόχρονα εξαρτάται από τα δείγματα όλων των προηγούμενων χρονικών βημάτων. \par 
\begin{figure}[!htb]
    \centering
    \includegraphics[scale=0.3]{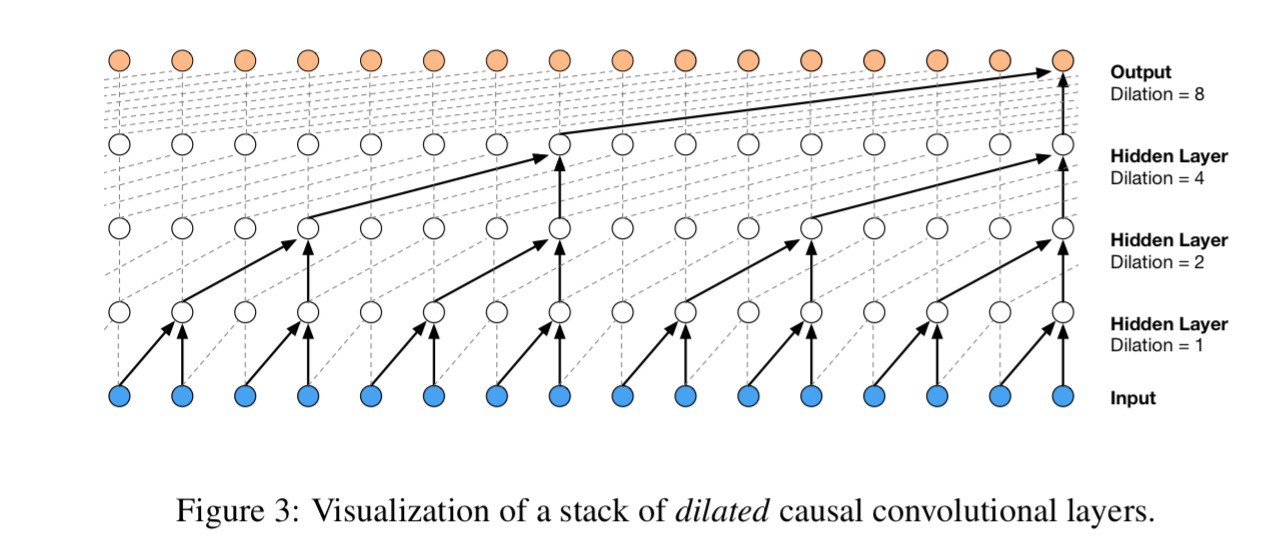}
        \caption{Dilated Convolutions from \cite{oord_wavenet_2016}}
    \label{fig:atrous}
\end{figure}
Προκειμένου να μοντελοποιηθούν οι πολύ μακροχρόνιες εξαρτήσεις (long-time dependencies) που εμφανίζονται στα ηχητικά σήματα, οι συγγραφείς χρησιμοποίησαν τις λεγόμενες διεσταλμένες συνελίξεις (dilated\footnote{Στη βιβλιογραφία αναφέρονται και ως a-trous που στη γαλλική σημαίνει ``με τρύπες".} convolutions).
Η βασική τους τροποποίηση έναντι του κλασικού CNN \cite{lecun_gradient-based_1998} είναι πως αυξάνουν το δεκτικό πεδίο (resceptive field) και συνεπώς τους δίνει τη δυνατότητα να σχετίζουν δείγματα της εισόδου τα οποία απέχουν πολύ στη χρονική διάσταση. Φυσικά όλα αυτά για να επιτευχθούν απαιτούν αρκετά επίπεδα αυξάνοντας κατά πολύ τις παραμέτρους του μοντέλου. Σχηματικά η βασική ιδέα φαίνεται και στο Σχήμα \ref{fig:atrous}.
\par 
Ο βασικός λόγος για τη δημοτικότητα του WaveNet είναι η παραγωγή φωνής υψηλής ποιότητας, της οποίας η φυσικότητα προσεγγίζει ανθρώπινα επίπεδα. Παρόλα αυτά το WaveNet πάσχει από δύο πολύ σοβαρά ζητήματα. Το πρώτο είναι η χρήση πολλαπλών συναρτήσεων κόστους (loss functions) κατά τη διάρκεια της εκπαίδευσης του δικτύου που μπορεί να οδηγήσει σε αστάθειες ή σε τετριμμένες λύσεις υψηλής εντροπίας (mode collapse) εαν δεν σταθμιστούν σωστά. Το δεύτερο και ίσως πιο ενδιαφέρον από πρακτική σκοπιά είναι η υπερβολικά αργή παραγωγή νέων δειγμάτων.
\par 

Εστιάζουμε στo δεύτερο μειονέκτημα, το οποίο είναι εγγενές χαρακτηριστικό της autoregressive φύσης του δικτύου. Μπορεί κατά τη διάρκεια της εκπαίδευσης το πρόβλημα αυτό να μην είναι εμφανές καθώς μας δίδεται η κυματομορφή εξόδου και συνεπώς όλοι οι υπολογισμοί μπορούν να παραλληλοποιηθούν, παρόλα αυτά κατά τη διάρκεια σύνθεσης θα πρέπει να περιμένουμε το δίκτυο να παράξει ένα δείγμα προκειμένου να παράξουμε το επόμενο. Συνεπώς η διαδικασία σύνθεσης δεν μπορεί να παραλληλοποιηθεί. Οι τρόποι που έχουνε προταθεί στη βιβλιογραφία για να αντιμετωπιστούν τα προβλήματα αργής σύνθεσης χωρίζονται σε μεθόδους που τείνουν να παραλληλοποιήσουν ή/και να απλοποιήσουν τους εμπλεκόμενους υπολογισμούς (engineering based acceleration) και σε μεθόδους που τείνουν να χρησιμοποιήσουν κάποια γνώση από πεδία όπως η επεξεργασία σήματος (knowledge-based acceleration).
\begin{figure}[!htb]
    \centering
    \includegraphics[scale=0.45]{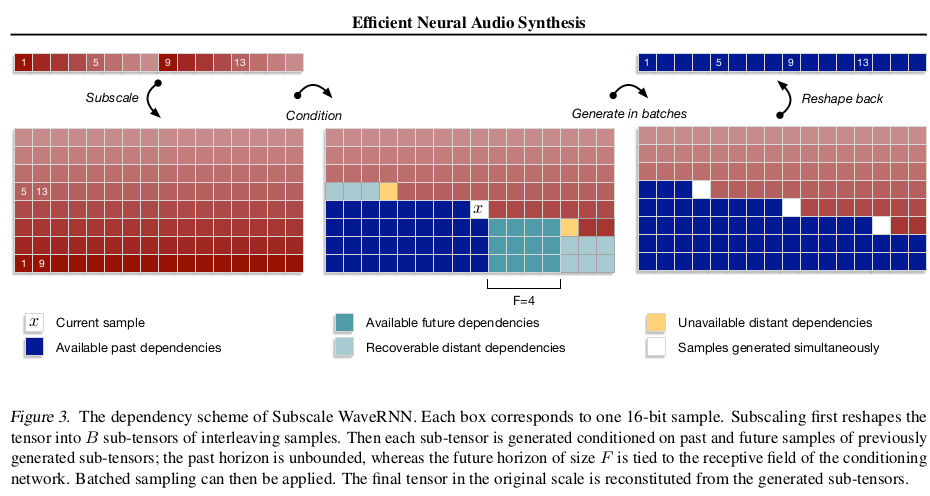}
    \caption{WaveRNN batch generation trick \cite{kalchbrenner_efficient_2018}}
    \label{fig:wave_rnn}
\end{figure}
\par
\vspace{5pt}
\noindent \textbf{Engineering Based Acceleration}. Σε αυτή την κατηγορία ανήκει το \textit{WaveRNN} \cite{kalchbrenner_efficient_2018} που είναι μια πιο αποδοτική προσέγγιση του αρχικού WaveNet. Η βασική ιδέα πίσω από αυτή την προσέγγιση είναι πως μπορούμε να παράξουμε πολλαπλά δείγματα την ίδια χρονική στιγμή. Άρα ανάλογα με το πόσα δείγματα επιλέγουμε να παράξουμε σε κάθε βήμα τόσο πιο γρήγορη γίνεται η προσέγγισή μας. Φυσικά υπάρχει ένα trade-off μεταξύ του πλήθους των παραγόμενων δειγμάτων και της τελικής ποιότητας. Σχηματικά η προσέγγιση όπως παρουσιάστηκε στην πρωτότυπη εργασία φαίνεται στο Σχήμα \ref{fig:wave_rnn}.

Μια ακόμη προσέγγιση που ονομάζεται \textit{FFTNet} \cite{jin_fftnet_2018} χρησιμοποιεί ένα trick που θυμίζει τον κλασικό αλγόριθμο FFT \cite{cooley_algorithm_1965} μιας και χωρίζει τα στοιχεία εισόδου σε δύο υποκατηγορίες που η κάθε μια περιέχει το μισό του αρχικού αριθμού δειγμάτων. Το ``σπάσιμο" αυτό μας επιτρέπει να υπολογίσουμε τις τελικές ποσότητες με παράλληλο τρόπο.
\par
\vspace{5pt}
\noindent \textbf{Knowledge Based Acceleration}. Μια υποσχόμενη προσέγγιση είναι το \textit{LPCNet} \cite{valin_lpcnet_2019} το οποίο, όπως υποδεικνύεται από το όνομά του, συνδυάζει τη γνωστή LPC μέθοδο από την επεξεργασία σημάτων \cite{o1988linear} με το WaveRNN, προκειμένου να αφαιρέσει ορισμένο υπολογιστικό φόρτο από την αρχική αρχιτεκτονική με απώτερο σκοπό να κάνει την εκπαίδευση ``ευκολότερη" και να δημιουργήσει γρηγορότερη παραγωγή νέων δειγμάτων.

\subsubsection{Normalizing Flow Approaches}
Αυτή η ενότητα εισάγει vocoders που εκμεταλλεύονται ένα ισχυρό στατιστικό εργαλείο, το οποίο χρησιμοποιείται για την εκτίμηση της πυκνότητας (density estimation) και ονομάζεται ομαλοποιμένη ροή (normalizing flow). Ένα γενετικό (generative) μοντέλο βασισμένο στη μέθοδο normalizing-flow κατασκευάζεται ως με μια ακολουθία διαδοχικών αναστρέψιμων μετασχηματισμών \cite{rezende2015variational, kingma_glow_2018}.
Σε γενικές γραμμές, αυτά τα μοντέλα μαθαίνουν ρητά (explicitly) την κατανομή δεδομένων $p(x)$ και επομένως η συνάρτηση κόστους (loss function) είναι απλώς η negative-log-likelihood.
\par 
\begin{figure}[!htb]
    \centering
    \includegraphics[scale=0.45]{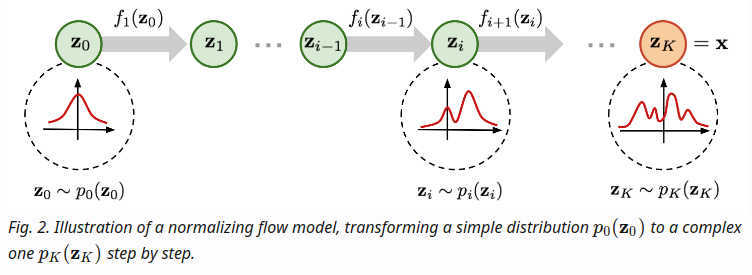}
    \caption{Normalizing Flow: Εκκινώντας από μια απλή κατανομή $p_0$ και εφαρμόζοντας διαδοχικούς μετασχηματισμούς οι οποίοι εκπαιδεύονται ($f_i$) καταλήγουμε σε μια προσέγγιση της υπό εξέτασης κατανομής $x$. (\href{https://lilianweng.github.io/lil-log/2018/10/13/flow-based-deep-generative-models.html}{Source})}
    \label{fig:nflow}
\end{figure}
\begin{figure}[!htb]
    \centering
    \includegraphics[scale=0.25]{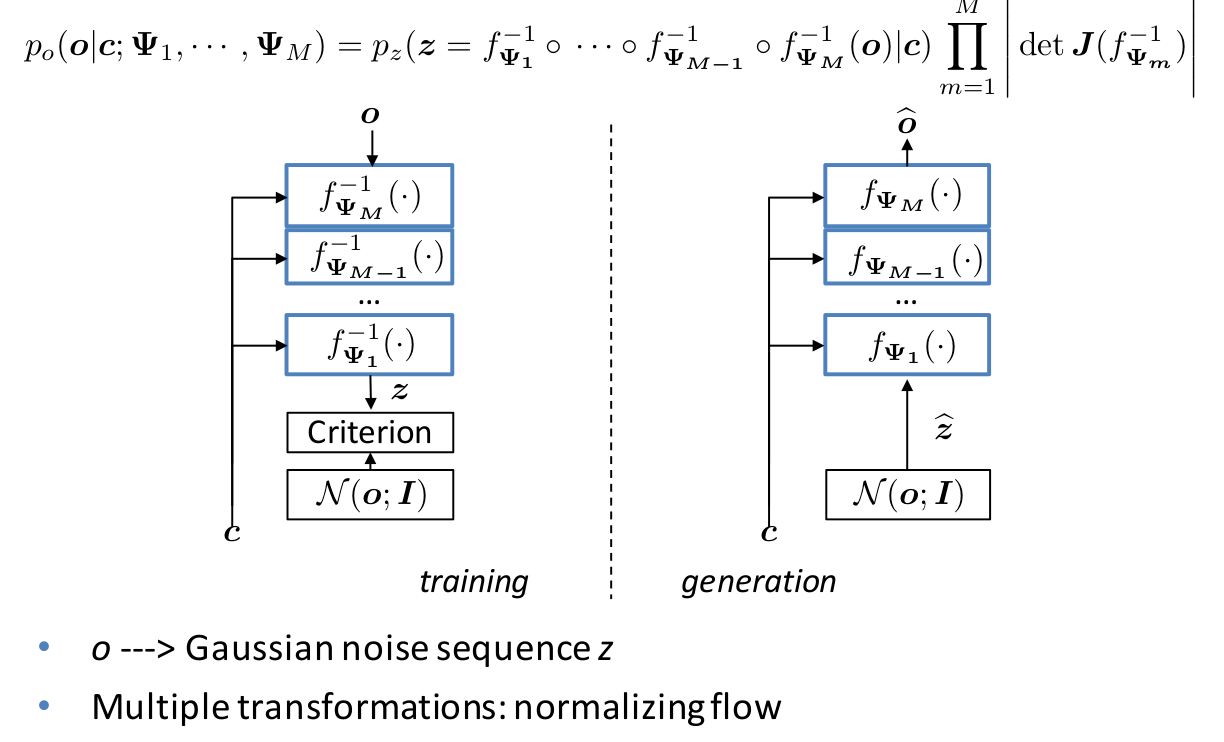}
    \caption{Normalizing Flow TTS (\href{https://www.slideshare.net/jyamagis/tutorial-on-endtoend-texttospeech-synthesis-part-1-neural-waveform-modeling}{Source})}
    \label{fig:tts_nflow}
\end{figure}

Στο πλαίσιο του προβλήματος σύνθεσης ομιλίας από κείμενο (TTS), η ιδέα των normalizing flows εισάγεται προκειμένου να ``απλοποιηθεί" η απεικόνιση που καλείται να μάθει το δίκτυο, όπως φαίνεται στο Σχήμα \ref{fig:tts_nflow}. 
Αυτό μπορούμε να το επιτύχουμε εφαρμόζοντας πολλαπλούς (αντίστροφους) μετασχηματισμούς ομαλοποίησης ροής (normalizing flow) στο one-hot διανύσματα της εξόδου $\mathbf{o}$. Η διαίσθηση πίσω από αυτό είναι ότι μετασχηματίζουμε οι ακολουθίες με ισχυρή χρονική συσχέτιση σε ακολουθίες με ασθενέστερους χρονικούς συσχετισμούς, απλοποιώντας έτσι την διαδικασία σύνθεσης.
Θα πρέπει να σημειωθεί ότι η επιλογή διαφορετικών επιμέρους μετασχηματισμών $f_{\Psi_m} (\cdot)$ αποφέρει φυσικά διαφορετικές χρονικές πολυπλοκότητες και επομένως διαφορετικές στρατηγικές εκπαίδευσης. Για να συνδέσουμε την παρούσα προσέγγιση με την αρχιτεκτονική WaveNet κάνουμε τις εξής δύο παρατηρήσεις. Το WaveNet είναι πλήρως παραλληλοποιήσμιο κατά τη διάρκεια της εκπαίδευσης (λόγω της συνελικτικής του φύσης) από τη στιγμή που δίνεται η κυματομορφή εξόδου. Αυτό δεν συμβαίνει με τη σύνθεση ωστόσο, όπου η πλήρης κυματομορφή δεν είναι διαθέσιμη και πρέπει να περιμένουμε να δημιουργηθεί ένα δείγμα για να προχωρήσουμε στο επόμενο.\par 
Τα μοντέλα \textit{Inverse Autoregressive Flow} (IAF) \cite{kingma2016improved} όπως αυτά που περιγράφονται σε αυτήν την ενότητα συμπεριφέρονται με δυϊκό τρόπο. Συγκεκριμένα παρουσιάζουν χαμηλότερη ταχύτητα εκαπίδευσης και υψηλότερη ταχύτητα σύνθεσης. Αυτό συμβαίνει διότι οι πολλαπλοί μετασχηματισμοί που εισάγονται αυξάνουν την πολυπλοκότητα του μοντέλου με αποτέλεσμα να είναι πιο δύσκολο για το μοντέλο να μάθει τους κατάλληλους normalizing flow μετασχηματισμούς. Εαν όμως οι ενδιάμεσοι μετασχηματισμοί ήταν γνωστοί με κάποιο τρόπο τότε η διαδικασία της σύνθεσης θα ήταν πολύ πιο απλή.
\begin{figure}[!htb]
    \centering
    \includegraphics[scale=0.3]{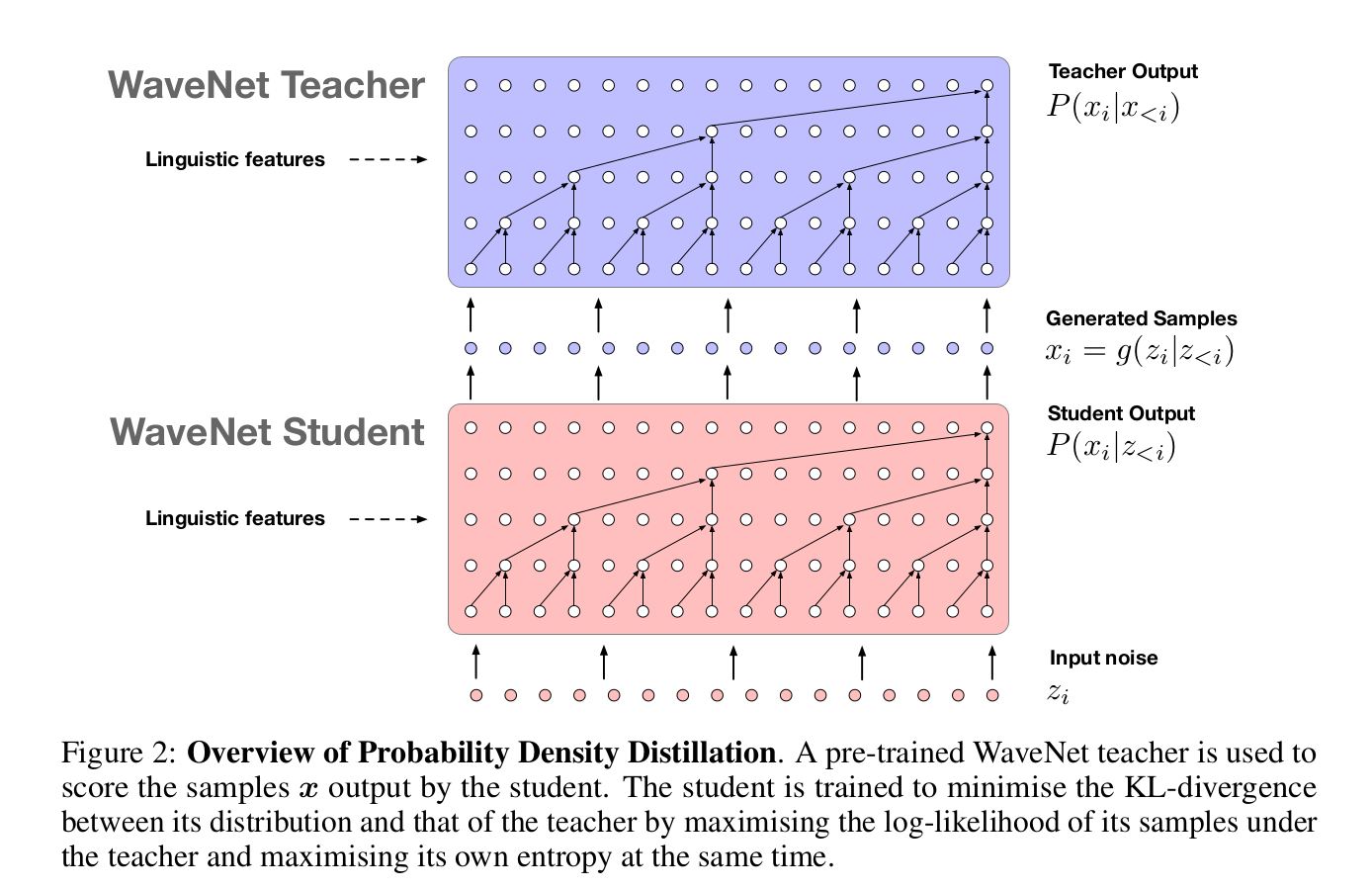}
    \caption{Parallel WaveNet \cite{oord_parallel_2017}}
    \label{fig:par_wavenet}
\end{figure}
\par 
Το \textit{ParallelWaveNet} \cite{oord_parallel_2017} είναι μια δουλειά από τους ίδιους συγγραφείς με το αρχικό WaveNet. Σε αυτήν εισάγεται η λεγόμενη \textit{Probability Density Distillation} που στοχεύει να συνδυάσει τα πλεονεκτήματα και από τα δύο μοντέλα. Με άλλα λόγια, δεδομένου ενός προ-εκπαιδευμένου WaveNet (parent), εκπαιδεύει ένα IAF WaveNet, όπως φαίνεται στο Σχήμα \ref{fig:par_wavenet}. H τεχνική knowledge distillation \cite{ba_deep_2014, hinton_distilling_2015} επιταχύνει τη διαδικασία εκπαίδευσης, καθώς οι ενδιάμεσες κατανομές δίνονται τώρα από το parent (γονικό) μοντέλο σε όλες τις ενδιάμεσες ροές. Αυτό έχει ως αποτέλεσμα την αρκετά ταχύτερη εκπαίδευση καθώς επίσης και την ταχύτερη σύνθεση φωνής.
\par
Μια ακόμη πιο αποτελεσματική προσέγγιση εισήχθη μέσω του λεγόμενου \textit{ClariNet} \cite{ping_clarinet_2019}, η οποία σε αντίθεση με τους συγγραφείς του Parallel WaveNet, χρησιμοποιεί αρκετά απλούστερους ενδιάμεσους μετασχηματισμούς και συγκεκριμένα διαδοχικούς Gaussian. H διαδικασία εκμάθησης των αντίστοιχων παραμέτρων έγκειται στην ελαχιτοποίηση μιας ομαλοποιημένης (regularized) KL-divergence (απόκλιση) μεταξύ των υψηλών κορυφών των κατανομών εξόδων τους.
\par
\vspace{5pt}
\noindent \textbf{Σύγκριση Parallel WaveNet και ClariNet}.
Παρόλο που και τα δύο δίκτυα συνθέτουν την ομιλία με παράλληλο τρόπο, χρειάζονται πολλαπλές συναρτήσεις κόστους (loss functions) για να δημιουργήσουν ρεαλιστική ομιλία \cite{oord_parallel_2017, ping_clarinet_2019} (η βιβλιογραφία αναφέρει το πρόβλημα ``ρομποτικής-μεταλλικής" φωνής ως το συνηθέστερο) και για να αποφύγουν προβλήματα mode collapse. Επιπλέον, το Parallel WaveNet χρησιμοποιεί μια Mixture of Logistics (MoL) κατανομή για τον προ-εκπαιδευμένο parent WaveNet. Η κατανομή αυτή όμως δεν διαθέτει λύση κλειστής μορφής για την απόκλιση KL και αντ'αυτού χρησιμοποιείται εκτίμηση τύπου Monte Carlo \cite{ceperley1977monte, doubilet1985probabilistic}. Η προσέγγιση Monte Carlo αναπόφευκτα εισάγει μια ανταλλαγή (trade-off) μεταξύ σταθερότητας και ταχύτητας. Το ClariNet, από την άλλη πλευρά, πρότεινε τη χρήση μιας και μόνο κατανομής Gauss, η KL απόκλιση της οποίας εκφράζεται με κλειστή μορφή και παρότι πιο απλή επιλύει μερικώς την αστάθεια που εισάγει η προσέγγιση του Monte Carlo που κάνει χρήση το Parallel WaveNet.
\par
\vspace{5pt}
\noindent \textbf{Επιλύοντας προβλήματα αστάθειας}.
Με σκοπό τώρα να αντιμετωπιστεί το πρόβλημα της αστάθειας κατά τη διάρκεια της εκπαίδευσης έχουν προταθεί κάποιες ακόμα μέθοδοι, οι οποίες επίσης βασίζονται στη μέθοδο normalizing flow. Συγκεκριμένα μοντέλα όπως το \textit{WaveGlow} \cite{prenger_waveglow_2018}, το \textit{FloWaveNet} \cite{kim_flowavenet_2019} και μετέπειτα το \textit{WaveFlow} \cite{ping2020waveflow}. Αυτά τα μοντέλα δημιουργούν ταχύτερη σύνθεση και εκπαίδευση σε επίπεδο ενός σταδίου, χρησιμοποιώντας μια μοναδική συνάρτηση κόστους. Η χρήση ενός μόνο loss function αντιμετωπίζει επιτυχώς τα ζητήματα σταθερότητας κατά τη διάρκεια εκπαίδευσης. Ωστόσο, λόγω της απουσίας ενός γονικού μοντέλου, η διαδικασία εκπαίδευσης διαρκεί περισσότερο από εκείνη των αντίστοιχων distillation προσεγγίσεων (Parallel WaveNet, ClariNet). Μια βασική ιδέα πίσω από αμφότερες τις προσεγγίσεις (WaveGlow, FloWaveNet) που βασίζονται σε normalizing flows είναι ο τελεστής squeeze. Στην ουσία η πράξη αυτή ομαδοποιεί τα ηχητικά δείγματα σε επιμέρους groups. Στο WaveGlow συγκεκριμένα ομαδοποιούνται τα δείγματα ανά $8$, ενώ στο FloWaveNet ομαδοποιούνται σε άρτια και περριτά. To WaveFlow με τη σειρά του παρέχει μια ενοποιημένη προσέγγιση η οποία περιέχει τόσο το WaveNet όσο και
το WaveGlow σαν ειδικές περιπτώσεις. Μελετά στην ουσία το trade-off μεταξύ παραλληλοποίησης και χωρητικότητας δικτύου. Σχηματικά παραθέτουμε το squeeze operation που υλοποιεί το WaveGlow καθώς επίσης και ένα δομικό στοιχείο της αρχιτεκτονικής του.

\begin{figure}[!htb]
    \centering
    \includegraphics[scale=0.25]{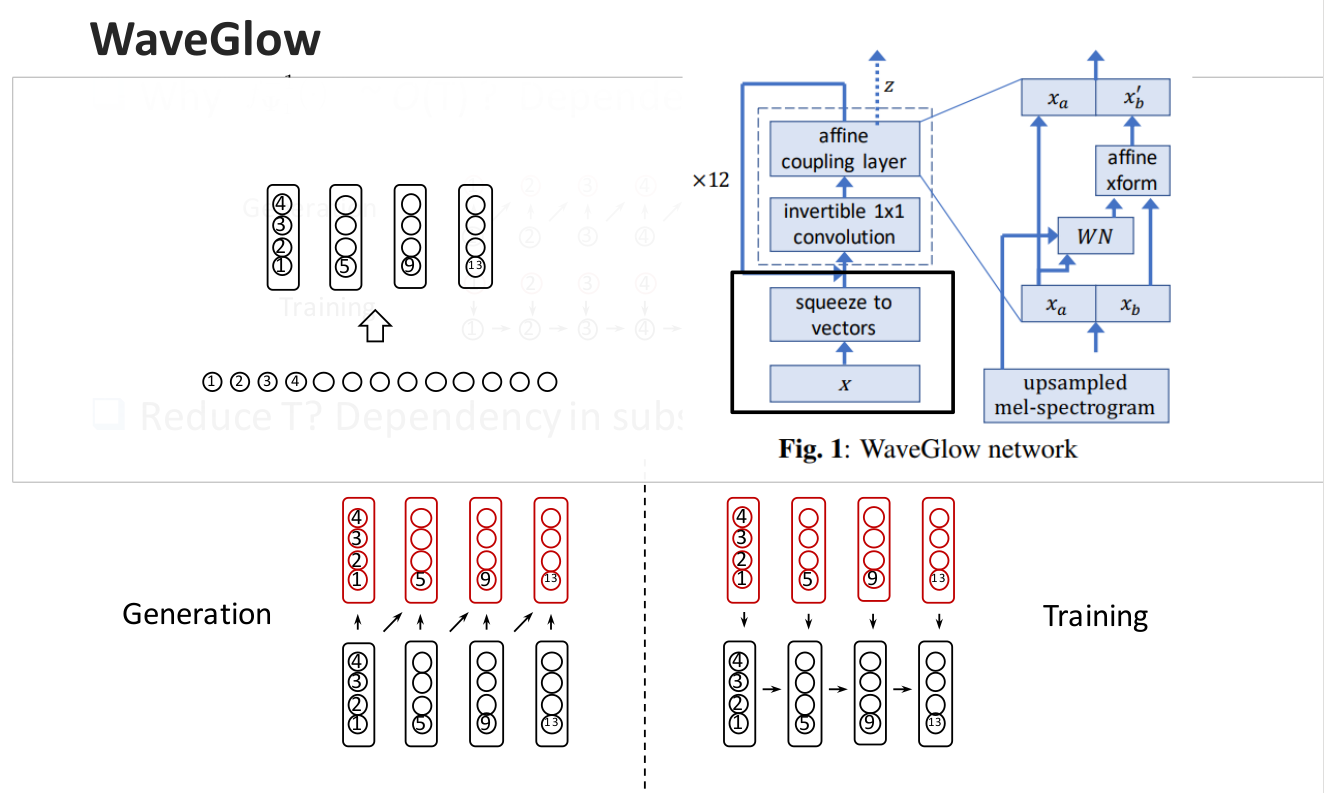}
    \caption{Wave Glow squeeze operation (\href{https://www.slideshare.net/jyamagis/tutorial-on-endtoend-texttospeech-synthesis-part-1-neural-waveform-modeling}{Source})}
    \label{fig:wave_glow}
\end{figure}

\subsubsection{GAN-based Προσεγγίσεις}
Στην ενότητα αυτή παρουσιάζονται εργασίες οι οποίες στηρίζονται στα λεγόμενα Generative Adversarial Networks (GANs) \cite{goodfellow_generative_2014} τα οποία εισήχθησαν το 2014 και έκτοτε έχουν βρει ευρύτατη εφαρμογή στην όραση υπολογιστών. \par 
Παρόλα αυτά, η εφαρμογή τους στο πεδίο της μοντελοποίησης ήχου/φωνής είναι μέχρι και σήμερα αρκετά περιορισμένη. Συγκεκριμένα προτάθηκε μια αρχιτεκτονική υπό το όνομα GANSynth \cite{engel_gansynth_2019}, η οποία χρησιμοποιείται για να παράγει με συνθετικό τρόπο διαφορετικά μουσικά ηχοχρώματα μοντελοποιοώντας STFT μέτρα και φάσεις\footnote{i.e magnitudes and angles of a complex number.}. Mια άλλη προσέγγιση \cite{neekhara_expediting_2019} απεικονίζει mel-specs σε απλά magnitude spectorgrams στα οποία μπορούν να εφαρμοστούν κατάλληλοι αλγόρθμοι ανάκτησης φάσης με απώτερο σκοπό την ανακατασκευή του αρχικού σήματος.
\par 
Μια εργασία που εφάπτεται στα αντικείμενα που πραγματεύεται η παρούσα βιβλιογραφική ανασκόπηση χρησιμοποιεί ένα GAN για να κάνει distill ένα autoregressive μοντέλο που παράγει συνθετική φωνή \cite{yamamoto_probability_2019}. Παρόλα αυτά τα πειραματικά τους αποτελέσματα έδειξαν πως χρησιμοποιώντας ένα adversarial loss από μόνο του δεν είναι αρκετό για να συνθέσει υψηλής ποιότητας κυματομορφές.
\par 
Η πιο σχετική εργασία \cite{kumar_melgan_2019} στη σύνθεση φωνής ονομάζεται \textit{MelGAN} και ουσιαστικά εισάγει μια non-autoregressive πλήρως συνελικτική προσέγγιση που παράγει κυματομορφές σε ένα GAN setup. Η προσέγγιση αυτή είναι αρκετά πιο γρήγορη στην παραγωγή φωνής απ'ότι οι αντίστοιχες autoregressive και περιέχει πολύ λιγότερες παραμέτρους.
\par 
Συγκεκριμένα γίνεται λόγος για $24.7$ m. παραμέτρους στο WaveNet \cite{shen_natural_2018}, για $10$ m. παραμέτρους στο ClariNet \cite{ping_clarinet_2019} και για $87.9$ m παραμέτρους στο WaveGlow \cite{prenger_waveglow_2018}, με το WaveGlow να δίνει τις καλύτερες επιδόσεις από άποψη ταχύτητας σύνθεσης. To \textit{MelGAN} έχει τόσο μικρότερο αριθμό παραμέτρων ($4.26$ m), όσο και πολύ υψηλότερη ταχύτητα παραγωγής κυματομορφών.
\par 
H αρχιτεκτονική του MelGAN αποτελείται προφανώς από έναν generator καθώς και έναν discriminator. Σε ότι αφορά το generator οι συγγραφείς αναφέρουν ότι τα επιμέρους μεγέθη της αρχιτεκτονικής έχουν επιλεγεί με πολύ προσεκτικό τρόπο διαφορετικά η προσέγγιση δεν ήταν επιτυχής. Για τον discriminator αναφέρεται πως χρησιμοποιείται μια multi-scale αρχιτεκτονική που περιέχει 3 discriminators, καθένας από τους οποίους αναλαμβάνει ήχο διαφορετικής κλίμακας (scale). Αναφέρουμε επίσης ότι τα τελικά αποτελέσματα παραγωγής φωνής κρίθηκαν υποδεέστερα άλλων προσεγγίσεων όπως το WaveGlow και το WaveNet.
\begin{figure}[!htb]
    \centering
    \includegraphics[scale=0.5]{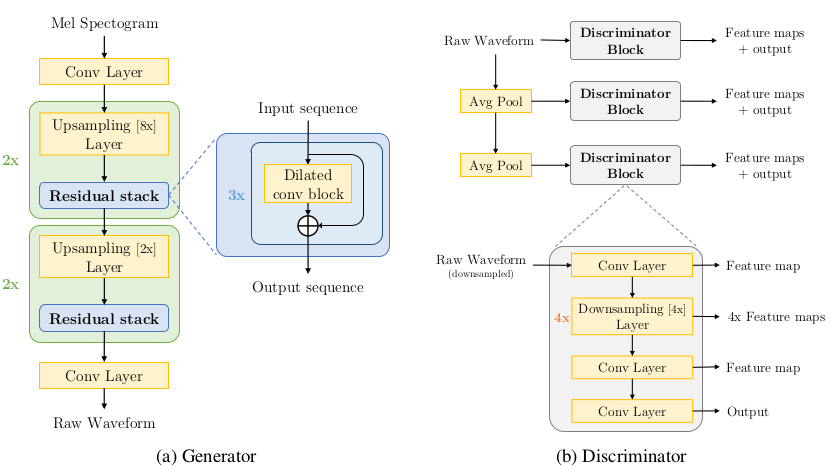}
    \caption{MelGAN \cite{kumar_melgan_2019}}
    \label{fig:melgan}
\end{figure}
\par 
\par
Mια ακόμα δουλειά είναι το \textit{TTS-GAN} \cite{binkowski_high_2019} που ουσιαστικά αποτελεί μια large-scale προσέγγιση του προβλήματος σύνθεσης φωνής. Η ουσιαστική συνεισφορά είναι πρακτικά η εκπαίδευση ενός μεγάλης κλίμακας GAN, με 44h δεδομένων και batch size της τάξης του 1024. Σε ότι αφορά το generator χρησιμοποιούνται 567 features ανα 5ms και σταδιακά εφαρμόζεται upsampling της αναπαράστασης, ενώ υπάρχουν παράλληλα και residual συνδέσεις. Συνολικά ο Generator περιέχει 30 επίπεδα. Σε ότι αφορά τον discriminator ουσιαστικά χρησιμοποιούνται πολλαπλοί discriminators που συγκρίνουν τυχαία τμήματα μιας κυματομορφής για διαφορετικά χρονικά παράθυρα (μεταβλητότητα ανάλυσης ως προς το χρόνο).

\subsection{Σύνοψη}
Στην ενότητα αυτή συνοψίζουμε τα βασικά χαρακτηριστικά των neural vocoders που παρουσιάστηκαν. Μια πρώτη κατηγοριοποίηση μπορεί να γίνει με βάση την αρχιτεκτονική προσέγγιση. Συγκεκριμένα μπορούμε να πούμε πως οι προσεγγίσεις χωρίζονται σε τρείς υποκατηγορίες i) τις autoregressive, και τις non-autoregressive που χωρίζονται στις ii) normalizing-flow και στις iii) GAN προσεγγίσεις. \par
Ως προς την ποιότητα της παραγόμενης συνθετικής φωνής το αρχικό WaveNet \cite{oord_wavenet_2016} είναι μέχρι και σήμερα η καλύτερη δυνατή προσέγγιση. Παρόλα αυτά πάσχει από δύο πολύ σοβαρά ζητήματα. Αφενός εξαιτίας της autoregressive φύσης και των πολλών παραμέτρων παρουσιάζει πολύ αργή σύνθεση φωνής, καθιστώντας την ακατάλληλη προσέγγιση για real-time εφαρμογές ενώ επιπλέον η εκπαίδευση είναι αρκετά επίπονη καθώς απαιτείται εκτεταμένη αναζήτηση κατάλληλων υπερπαραμέτρων ώστε η προσέγγιση να συγκλίνει τελικά. \par 
Για τους παραπάνω λόγους έχουν προταθεί τόσο οι προσεγγίσεις normalizing flow όπως και οι generative adversarial. Σε ότι αφορά τις normalizing flow δουλειές, οι περισσότερες χρειάζονται ένα προεκαπιδευμένο γονέα WaveNet πράγμα που αυτόματα συνεπάγεται πως πρέπει να εκπαιδεύσουμε ένα τέτοιο μοντέλο πρώτα. Το πρόβλημα που αντιμετωπίζουν οι προσεγγίσεις αυτές είναι η αργή σύνθεση φωνής. Το parent μοντέλο χρειάζεται καθώς οι πρσεγγίσεις αυτές αυξάνουν τον αριθμό των παραμέτρων καθιστώντας σχεδόν αδύνατο να εκπαιδευτεί ένα τόσο ογκώδες δίκτυο και να πετύχει αποτελέσματα κοντά στα ανθρώπινα επίπεδα. 
\par 
Η πιο σημαντική ίσως προσέγγιση που επιλύει και τα δύο προαναφερθέντα προβλήματα ως ένα βαθμό χωρίς όμως να μειώνει πολύ την ποιότητα της παραγώμενης φωνής είναι το WaveGlow \cite{prenger_waveglow_2018} το οποίο χρησιμοποιεί ένα trick παραγωγής πολλών δειγμάτων ταυτόχρονα και επιπλέον χρησιμοποιεί ένα και μόνο loss κάνοντας την εκπαίδευσή του τετριμμένη διαδικασία. \par 
Τέλος, το MelGAN \cite{kumar_melgan_2019} είναι μια προσέγγιση η οποία χρησιμοποιεί GANs και είναι η πρώτη που πετυχαίνει καλά αποτελέσματα ως προς την ποιότητα της παραγόμενης φωνής, παρόλα αυτά υποδεέστερα των υπόλοιπων προσεγγίσεων. Στα θετικά της μεθόδου επίσης συγκαταλέγεται ο χαμηλός αριθμός παραμέτρων, ενώ στα αρνητικά η πολύ προσεκτικά επιλεγμένη αρχιτεκτονική του generator.

\section{Απεικόνιση κειμένου σε φασματογράφημα και end-to-end προσεγγίσεις στη σύνθεση φωνής}
Η τεχνητή παραγωγή ανθρώπινης ομιλίας είναι γνωστή ως σύνθεση ομιλίας. Αυτή η τεχνική που βασίζεται στη μηχανική εκμάθηση μπορεί να εφαρμοστεί σε πληθώρα εφαρμογών όπως κείμενο-σε-ομιλία, δημιουργία μουσικής, παραγωγή ομιλίας, συσκευές με δυνατότητα ομιλίας, συστήματα πλοήγησης και προσβασιμότητα για άτομα με προβλήματα όρασης. Αλλά πριν προχωρήσουμε, υπάρχουν μερικές συγκεκριμένες, παραδοσιακές στρατηγικές για τη σύνθεση ομιλίας που πρέπει να περιγράψουμε εν συντομία: συναθροιστικές (concatenative) και παραμετρικές (parametric).
\par
Στην concatenative προσέγγιση, οι ομιλίες από μια μεγάλη βάση δεδομένων χρησιμοποιούνται για τη δημιουργία νέας, ακουστικής ομιλίας. Σε περίπτωση που απαιτείται διαφορετικό στυλ ομιλίας, χρησιμοποιείται μια νέα βάση δεδομένων με φωνητικές φωνές. Αυτό περιορίζει την επεκτασιμότητα αυτής της προσέγγισης.
Η παραμετρική προσέγγιση χρησιμοποιεί μια καταγεγραμμένη ανθρώπινη φωνή και μια συνάρτηση με ένα σύνολο παραμέτρων που μπορούν να τροποποιηθούν για να αλλάξουν τη φωνή. \textbf{Ωστόσο, οι προσεγγίσεις Deep Learning δημιουργούν μια ακολουθία νευρωνικών μοντέλων που στοχεύουν στην απεικόνιση της ακολουθίας κειμένου στην πρωτογενή κυματομορφή ήχου.}
\par
Στην ενότητα αυτή θα μελετήσουμε προσεγγίσεις που βασίζονται κυρίως στην κατασκευή του ``front-end" στη σύνθεση φωνής. Δηλαδή ρχιτεκτονικές που αναλαμβάνουν να απεικονίσουν μια λεκτική περιγραφή σε κάποια ενδιάμεση ακουστική αναπαράσταση από την οποία μετά θα συντεθεί το τελικό ηχητικό σήμα.

\subsection{Προσεγγίσεις στη βιβλιογραφία}
\begin{figure}[!htb]
    \centering
    \includegraphics[scale=0.5]{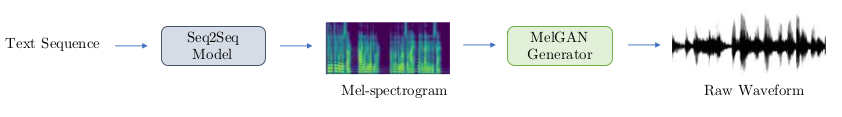}
    \caption{End-to-end pipeline from \cite{kumar_melgan_2019}}
    \label{fig:e2e_pipeline}
\end{figure}

Στην ενότητα αυτή θα παρουσιάσουμε κάποιες δουλειές που εστιάζουν είτε στο πρώτο κομμάτι της παραγωγής spectrogram (seq2seq του Σχήματος \ref{fig:e2e_pipeline}) από κείμενο είτε στην πλήρη end-to-end προσέγγιση.
Σχηματικά η διαδικασία φαίνεται παρακάτω.
Να σημειωθεί πως το σχήμα αναφέρεται στην παραγωγή mel-spectrogram, ενώ το πράσινο πλαίσιο που αναγράφει MelGAN Generator μπορεί να αντικατσταθεί από οποιαδήπποτε neural vocoder αρχιτεκτονική.
\par
Μια προσέγγιση στη σύνθεση φωνής από κείμενο αποτελεί η ``τριλογία" της Baidu με όνομα \textit{DeepVoice} 1 \cite{arik_deep_2017}, DeepVoice 2 \cite{arik_deep_2017-1} και DeepVoice 3 \cite{ping_deep_2018}. Στην τρίτη προσέγγιση που είναι και η πιο ολοκληρωμένη οι συγγραφείς εισήγαγαν μια πλήρως συνελικτική (fully convolutional) αρχιτεκτονική η οποία επιπλέον είχε και μηχανισμούς προσοχής (attention mechanism). Σκοπός της αρχιτεκτονικής αυτής είναι να απεικονίσει textual features όπως characters, phonemes και stresses σε διαφορετικές παραμέτρους ενός vocoder όπως mel-band spectrograms, linear-scale log magnitude spectrograms, θεμελιώδης συχνότητα ($F_0$), spectral envelope και παραμέτρους απεριοδικότητας (aperiodicity). Αυτές οι παράμετροι χρησιμοποιούνται στη συνέχεια ως είσοδοι για τον vocoder. H συνολική αρχιτεκτονική έχει τρία μέρη και συγκεκριμένα έναν encoder που παίρνει τα textual features σε μια ``εσωτερική" απεικόνιση, έναν decoder που αποκωδικοποιεί τις ενδιάμεσες αυτές αναπαραστάσεις και τέλος έναν converter που προβλέπει τις τελικές παραμέτρους του vocoder.
\par 
Μια άλλη προσέγγιση με αρκετά πιο απλό αρχιτεκτονικό σχεδιασμό είναι το \textit{Char2Wav} \cite{Sotelo2017Char2WavES}. To Char2Wav, είναι ένα end-to-end μοντέλο για σύνθεση ομιλίας και αποτελείται από δύο επιμέρους τμήματα. Έναν reader και έναν neural vocoder. Ο reader είναι ένα μοντέλο encoder-decoder με attention. Ο encoder είναι ένα αμφίδρομο (bidirectional) ακολουθιακό νευρικό δίκτυο (RNN) που δέχεται κείμενο ή φωνήματα ως εισόδους, ενώ ο decoder είναι επίσης ένα RNN με attention που παράγει ακουστικά χαρακτηριστικά για τον vocoder. Ως vocoder προτείνεται μια conditional επέκταση του SampleRNN \cite{mehri2016samplernn} που παράγει δείγματα κυματομορφής από ενδιάμεσες αναπαραστάσεις. Σε αντίθεση με τα παραδοσιακά μοντέλα για τη σύνθεση ομιλίας, το Char2Wav μαθαίνει να παράγει ήχο απευθείας από το κείμενο.\par 
Από την ομάδα Facebook AI Research \cite{taigman_voiceloop_2018} παρουσιάστηκε το \textit{VoiceLoop}, μια προσέγγιση που μετασχηματίζει το κείμενο σε φωνές οι οποίες έχουν ληφθεί in the wild\footnote{Σε ελεύθερη απόδοση σημαίνει σε μη ιδανικές-εργαστηριακές συνθήκες.}. Η προσέγγιση βασίζεται σε ένα μοντέλο λειτουργικής μνήμης γνωστό ως φωνολογικό βρόχο, το οποίο κρατά λεκτικές πληροφορίες για μικρό χρονικό διάστημα. Αποτελείται από ένα φωνολογικό κελί μνήμης, που αντικαθίσταται συνεχώς και μια διαδικασία, η οποία διατηρεί μακροπρόθεσμες (long-term) αναπαραστάσεις στη φωνολογική μνήμη. Το VoiceLoop κατασκευάζει τη φωνολογική μνήμη εφαρμόζοντας ένα buffer ολίσθησης ως μήτρα. Οι προτάσεις αναπρίστανται ως λίστα φωνημάτων. Στη συνέχεια, ένα διάνυσμα χαμηλής διάστασης αποκωδικοποιείται από κάθε ένα από τα φωνήματα. Το τρέχον context vector δημιουργείται σταθμίζοντας την κωδικοποίηση των φωνημάτων και αθροίζοντας τα σε κάθε χρονικό σημείο. Αυτό που διαφοροποιεί το VoiceLoop είναι η χρήση ενός buffer μνήμης αντί των συμβατικών RNN, η κοινή χρήση μνήμης μεταξύ όλων των διαδικασιών και χρήση shallow, πλήρως συνδεδεμένων δικτύων για όλους τους υπολογισμούς.
\par
Στις end-to-end προσεγγίσεις κατατάσσεται και η πρόσφατη δουλειά της DeepMind \cite{donahue2020endtoend}. Ο προτεινόμενος generator είναι εμπρόσθιας τροφοδότησης κάνοντας το έτσι αποτελεσματικό τόσο στη διάρκεια εκπαίδευσης όσο και κατά τη σύνθεση. Επιπλέον χρησιμοποιεί ένα διαφορίσιμο alignment σχήμα, το οποίο βασίζεται στην πρόβλεψη της διάρκειας κάθε token. Το συνολικό δίκτυο μαθαίνει μέσω ενός συνδυασμού από adversarial feedback και πολλαπλά losses αναγκάζοντας την τελική κυματομορφή να μοιάζει με την αρχική σε συνολική διάρκεια και mel-spectrogram. Επιλέον ένα soft dynamic time warping εφαρμόζεται ώστε το δίκτυο να μαθαίνει τοπικές χρονικές μεταβλητότητες.
\par
Αρκετές ακόμα προσεγγίσεις υπάρχουν στη βιβλιογραφία παρόλα αυτά θα επικεντρωθούμε σε αυτές που δίνουν τα καλύτερα αποτελέσματα και σε ορισμένες άλλες ανερχόμενες.
\subsection{Tacotron \& Tacotron 2}
\begin{figure}[!htb]
    \centering
    \includegraphics[scale=0.3]{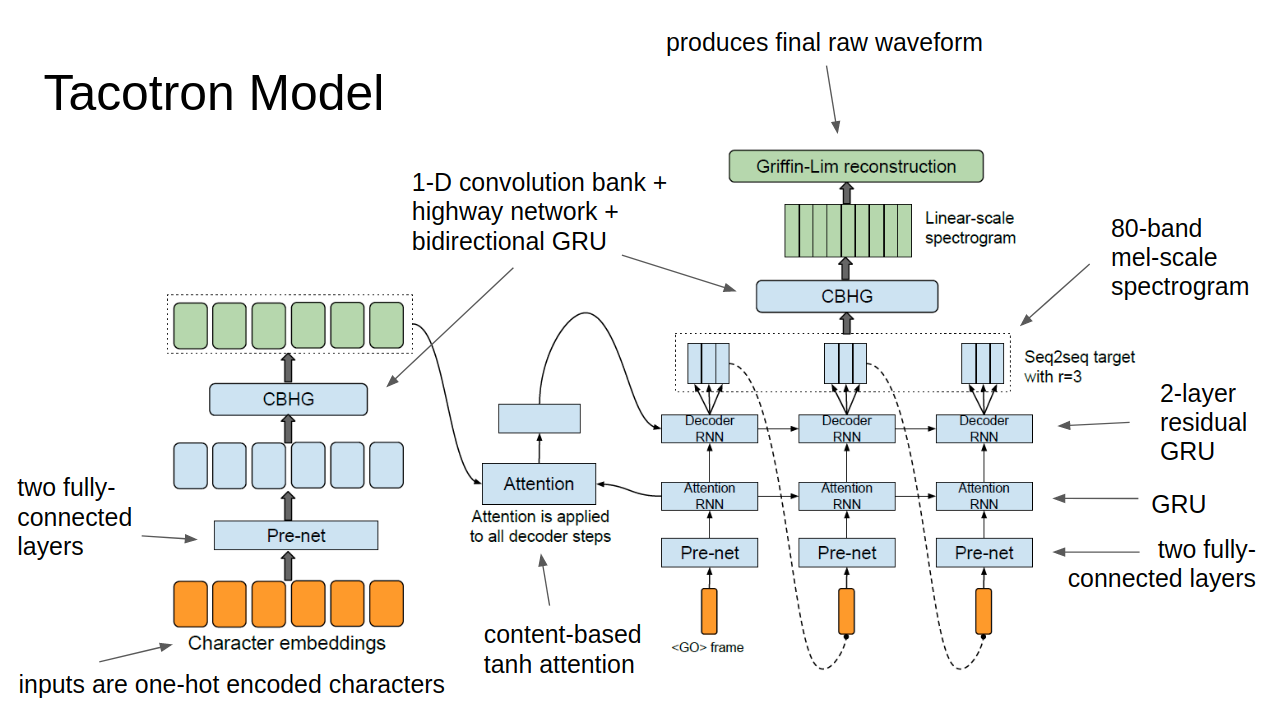}
    \caption{Tacotron \cite{wang_tacotron_2017}}
    \label{fig:tacotron}
\end{figure}

Και οι δύο δουλειές που αναφέρονται ως Tacotron αι Tacotron 2 αποτελούν έρευνα της Google πάνω στο πρόβλημα της απευθείας συνθεσης φωνής από κείμενο. Το Tacotron είναι το πρώτο νευρωνικό δίκτυο που πέτυχε καλύτερα αποτελέσματα από τις παραδοσιακές συνδυαστικές (concatenative) και παραμετρικές (parametric) προσεγγίσεις στη σύνθεση ομιλίας. Η αρχιτεκτονική του απεικονίζεται παρακάτω στο σχήμα \ref{fig:tacotron}. Τα κύρια πλεονεκτήματά του είναι ότι μπορεί να εκπαιδευτεί σε ζεύγη ήχου κειμένου που το καθιστά πραγματικά κατάλληλο για πολλά υπάρχοντα σύνολα δεδομένων (είναι σύνηθες να εκπαιδεύονται νευρωνικά δίκτυα σε δεδομένα τα οποία στη βιβλιογραφία έχουν χρησιμοποιηθεί για ASR στο παρεθλόν).

Αναλύοντας προσεκτικά την αρχιτεκτονική του Tacotron \cite{wang_tacotron_2017} μπορούμε να παρατηρήσουμε τα ακόλουθα αρχιτεκτονικά μπλοκ. Πρώτα απ 'όλα είναι ένα μοντέλο sequence-to-sequence που απεικονίζει one-hot encodings, επίπεδου χαρακτήρων, σε φασματογραφήματα (spectrograms). Συγκεκριμένα μια ακολουθία χαρακτήρων τροφοδοτείται στον encoder που εξάγει διαδοχικές αναπαραστάσεις κειμένου. Ένας αποκωδικοποιητής (decoder) που βασίζεται σε μηχανισμό προσοχής είναι τότε υπεύθυνος για το ``δύσκολο" έργο της απεικόνισης σε φασματογράφημα. Τέλος το φασματογράφημα που παράγεται συντίθεται σε (συνθετική) φωνή μέσω του αλγορίθμου Griffin-Lim hl{add citation}.
\par
Το Tacotron 2 \cite{shen_natural_2018} είναι ένα πλήρως νευρικό μοντέλο TTS. Αποτελείται κυρίως από δύο μέρη και συγκεκριμένα από ένα μοντέλο seq2seq, το οποίο απεικονίζει χαρακτήρες σε mel-specs καθώς και ένα τροποποιημένο WaveNet Vocoder. Η πλήρης αρχιτεκτονική είναι πιο απλή από αυτήν που χρησιμοποιείται στο Tacotron, δεδομένου ότι χρησιμοποιούνται μόνο κλασικά LSTM \cite{hochreiter1997long} και συνελίξεις. Παρόλα αυτά η πλήρης αρχιτεκτονική παραμένει αρκετά περίπλοκη για να εξηγηθεί λεπτομερώς εδώ.\par

\begin{figure}[!htb]
    \centering
    \includegraphics[scale=0.3]{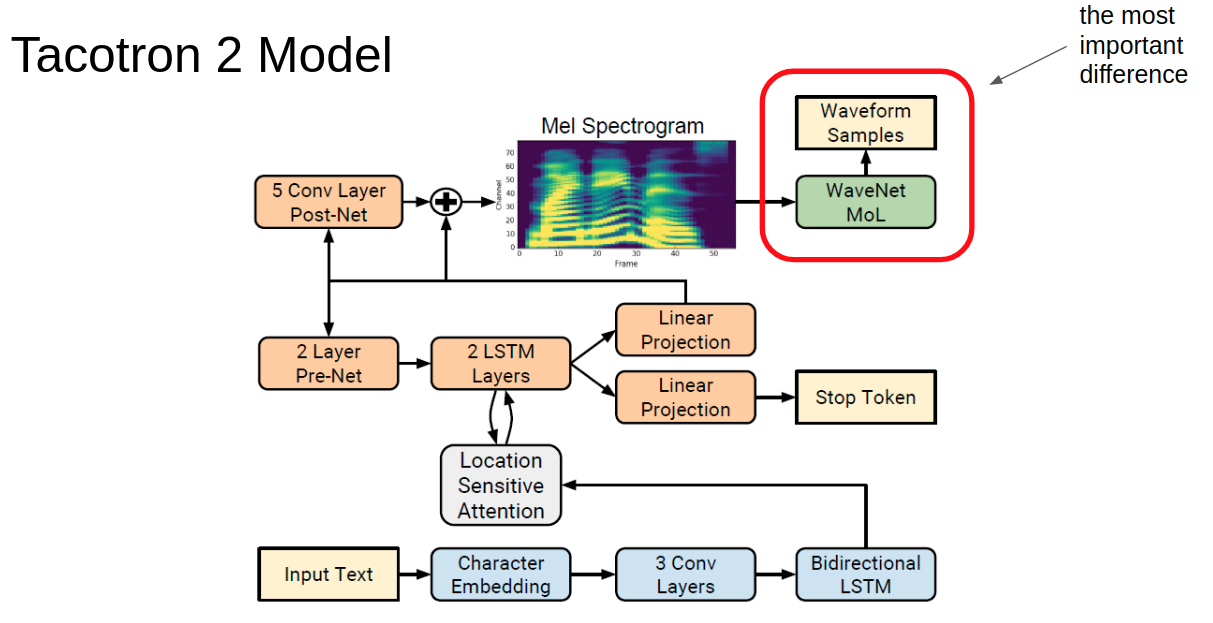}
    \caption{Tacotron 2 \cite{shen_natural_2018}}
    \label{fig:tacotron_2}
\end{figure}

To πιο σημαντικό μέρος του δικτύου, σύμφωνα με τους συγγραφείς της εργασίας, είναι το τροποποιημένο WaveNet. Με τον όρο τροποποίηση οι συγγραφείς σημαίνουν όχι μόνο τις αρχιτεκτονικές διαφορές (λιγότερες παραμέτρους από το πρωτότυπο WaveNet), αλλά κυρίως την προσαρμογή του σε χαρακτηριστικά mel. Η κύρια δουλειά που έχει τώρα ο vocoder είναι να αναστρέψει τo mel spectrogram. \par
Όσον αφορά το seq2seq τμήμα, το αθροιστικό MSE (μέσο τετραγωνικό) σφάλμα χρησιμοποιείται πριν και μετά το post-net ώστε να βοηθήσει τη σύγκλιση του δικτύου. Yπάρχει ένα επιπλέον stop token, το οποίο επιτρέπει στο μοντέλο να προσδιορίζει δυναμικά πότε να τερματίζει τη δημιουργία αντί να δημιουργεί πάντα ακολουθίες μιας σταθερής διάρκειας. \par
Η διαδικασία εκπαίδευσης χωρίζεται στη συνέχεια σε δύο μέρη. Πρώτα απ 'όλα, το μοντέλο seq2seq εκπαιδεύεται για να αντιστοιχίσει τα χαρακτηριστικά από το χώρο των χαρακτήρων σε mel-spectrograms. Οι συγγραφείς ισχυρίζονται ότι μπορούν να εκπαιδεύσουν ένα τέτοιο δίκτυο σε μία μόνο GPU. \par 
Όσο για το τροποποιημένο WaveNet εκπαιδεύεται στα πραγματικά συγχρονισμένα φασματογραφήματα (\textit{ground true aligned predictions}) του δικτύου προβλέψεων spectrogram. Με άλλα λόγια, χρησιμοποιούνται τα φασματογραφήματα που παράγει το αρχικό δίκτυο αλλά με μια δίορθωση ως προς το συγχρονισμό (alignment) τους. Οι συγγραφείς ισχυρίζονται ότι \textit{αυτή η μειωμένη έκδοση WaveNet απαιτούσε 32 GPU για εκπαίδευση.} 
\par
Όσον αφορά τις ενδιάμεσες αναπαραστάσεις (spectrograms), εκτενής πειραματισμός με μια ποικιλία από αυτές δείχνει την υπεροχή των φασματογραφημάτων mel. Μια ενδιαφέρουσα πτυχή είναι επίσης ότι το τμήμα vocoder εκπαιδεύτηκε επίσης σε mel-specs που εξήχθησαν από ground truth ήχο. Αυτό είχε ως αποτέλεσμα χειρότερη απόδοση από εκείνη του αρχικού μοντέλου. 
\par
Τέλος, χρησιμοποιήθηκε μια πιο ελαφριά έκδοση του WaveNet, την οποία οι συγγραφείς πιστεύουν ότι κατάφεραν να ενσωματώσουν στην αρχιτεκτονική λόγω της χρήσης των φασματογραφημάτων mel-spec, καθώς τα mel-specs μπορούν να συλλάβουν μακροπρόθεσμες εξαρτήσεις ήχου. Ωστόσο, εξακολουθούν να απαιτούνται διασταλτικές συνελίξεις (dilated convolutions), δεδομένου ότι παρέχουν στο μοντέλο επαρκές πλαίσιο στη χρονική κλίμακα των δειγμάτων κυματομορφής.

\subsection{Transformer-based approaches}
H επεξεργασία ακολουθιακών δεδομένων όπως το κείμενο και ο ήχος έχουν απασχολήσει αρκετά την επιστημονική κοινότητα. Η παραδοσιακή προσέγγιση αφορά τα λεγόμενα RNN, η αιχμή του δόρατος των οποίων είναι το διάσημο LSTM, τα οποία είναι μια προσπάθεια να επεξεργαστούμε ακολουθίες μεταβλητού μήκους. Μια εναλλακτική προσέγγιση που είδαμε στη μέχρι τώρα ανασκόπηση είναι η χρήση stacked dilated convolutional layers τα οποία επιτρέπουν στο δίκτυο να μαθαίνει χρονικές εξαρτήσεις σε μεγάλα χρονικά πλαίσια.
\par 
Μια επιπλέον προσέγγιση είναι ο λεγόμενος \textit{Transformer} \cite{vaswani2017attention}, ο οποίος ουσιαστικά επεργάζεται ακολουθίες βασιζόμενες σε (πυνκούς) μηχανισμούς προσοχής. Χρησιμοποιώντας πολλαπλές κεφαλές προσοχής (attention heads) μαθαίνει διαφορετικές ``υφές" της ακολουθίας. Η χρήση τους στην Επεξεργασία Φυσικής Γλώσσας (NLP) έχει δώσει τα μέχρι και σήμερα καλύτερα αποτελέσματα (tate-of-the-art) σε πολλά tasks.
\par 
Στην ενότητα αυτή εξετάζουμε πώς έχει χρησιμοποιηθεί αυτή η αρχιτεκτονική στη σύνθεση ομιλίας. Τα δύο βασικά μειονεκτήματα της προσέγγισης του Tacotron 2 \cite{shen_natural_2018} είναι i) η χαμηλή αποδοτικότητα κατά τη διάρκεια της εκπαίδευσης και σύνθεσης, ii) η μη χρήση ακολουθιακών νευρωνικών δικτύων σε όλη την έκταση της αραχιτεκτονικής.

\begin{figure}[!htb]
    \centering
    \includegraphics[scale=0.4]{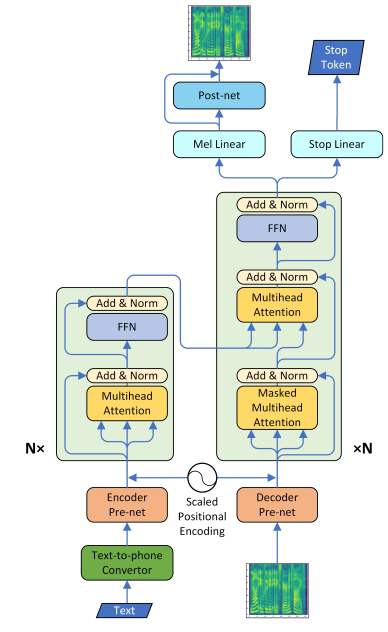}
    \caption{Transformer Based TTS \cite{li_neural_2019}}
    \label{fig:transformer_tts}
\end{figure}

Στο πλαίσιο της σύνθεσης φωνής από κείμενο, ερευνητές από τη Microsoft \cite{li_neural_2019} ήταν οι πρώτοι που αντικατέστησαν την αρχιτεκτονική του Tacotron με μια βασισμένη σε Transformer αρχιτεκτονική. Το τμήμα του vocoder εξακολουθεί να είναι αρχιτεκτονική που βασίζεται στο WaveNet. Ο μηχανισμός self-attention πολλαπλών κεφαλών (multihead) λύνει αποτελεσματικά το πρόβλημα μακροπρόθεσμης εξάρτησης σε ακολουθιακά δεδομένα. Αν και αυτή η προσέγγιση αυξάνει την ταχύτητα εκπαίδευσης (λόγω πλήρους παραλληλοποίησης), προκύπτει ένα επιπλέον ζήτημα που είναι ο διπλασιασμός των εκπαιδευόμενων παραμέτρων. Επιπλέον, η εξάρτηση από προηγούμενες τιμές κατά τη σύνθεση δεν άρεται και συνεπώς ούτε και η σχετική επιβράδυνση. Η προτεινόμενη αρχιτεκτονική απεικονίζεται και στο Σχήμα \ref{fig:transformer_tts} .
\par
Σε μια προσπάθεια αντιμετώπισης των προαναφερθέντων ζητημάτων, καθώς και προς την επίτευξη ευρωστίας (η επανάληψη ορισμένων λέξεων είναι σύνηθες σε μεγάλες προτάσεις), δημοσιεύτηκε μια εργασία με το όνομα \textit{FastSpeech} \cite{ren_fastspeech_2019}. Όπως αναφέρεται στο πρωτότυπο κείμενο ``\textit{Λαμβάνοντας υπόψιν τη μονότονη ευθυγράμμιση (alignment) μεταξύ κειμένου και ομιλίας, και με απώτερο σκοπό την επιτάχυνση της σύνθεσης mel specotrograms, προτείνουμε ένα νέο μοντέλο (FastSpeech) το οποίο παίρνει ένα κείμενο, ακολουθία φωνημάτων,
ως είσοδο και παράγει mel-φασματογραφήματα non-autoregressively\footnote{Δηλαδή κατά τη σύνθεση δεν χρειάζεται να περιμένουμε να παράξουμε ένα δείγμα προκειμένου να παραχθεί κάποιο μεταγενέστρο του χρονικά}.}"
\begin{figure}[!htb]
    \centering
    \includegraphics[scale=0.4]{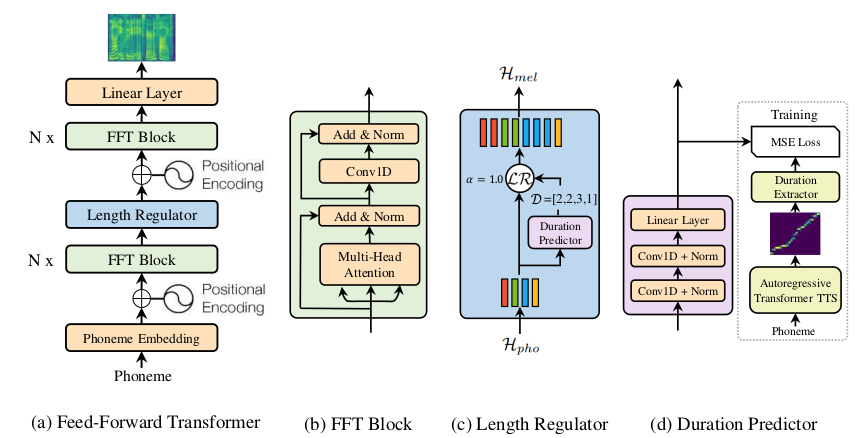}
    \caption{FastSpeech \cite{ren_fastspeech_2019}}
    \label{fig:fastspeech}
\end{figure}
\par
Συγκεκριμένα χρησιμοποιείται ένα δίκτυο εμπρόσθιας τροφοδότησης βασισμένο σε μηχανισμό multihead self-attention και σε μονοδιάστατες (1D) συνελίξεις. Εξαιτίας της μη-συμβατότητας μεταξύ των μηκών της ακολουθίας φωνημάτων και του mel-spectrogram, χρησιμοποιείται ένας length regulator που δειγματοληπτεί την ακολυθία φωνημάτων σύμφωνα με τη διάρκεια του κάθε φωνήματος, ώστε τελικά να ταιριάζουν τα δύο μήκη. Ο ρυθμιστής αυτός προφανώς βασίζεται σε ένα σύστημα πρόβλεψης διάρκειας φωνήματος.
\par
Αν και το FastSpeech επιλύει αποτελεσματικά πολλά προβλήματα των αρχικών αρχιτεκτονικών που βασίζονται σε Transformer, απαιτεί ακόμα σημαντικό αριθμό GPU για να εκαπιδευτεί, ενώ επιπλέον χρειάζεται και ένα γονικό (parent) προ-εκπαιδευμένο δίκτυο WaveNet που ουσιαστικά χωρίζει την εκπαίδευση σε δύο στάδια.

\subsection{Σύνοψη end-to-end μεθόδων}
Στην ενότητα αυτή συνοψίζουμε τα θετικά και τα αρνητικά των προσεγγίσεων που παρουσιάσαμε. Συγκεκριμένα θα πρέπει να ανφερθεί πως καθαρά από άποψη επίδοσης το Tacotron 2 πετυχαίνει τα καλύτερα δυνατά αποτελέσματα όπως κρίθηκαν από ανθρώπους. Στη συνέχεια ο Transformer TTS επιτυγχάνει εξίσου καλά αποτελέσματα με το Tacotron 2 αλλά αυξάνει δραματικά τον αριθμό των παραμέτρων και είναι αργός στη διαδικασία σύνθεσης. Τέλος το FastSpeech φαίνεται σαν μια υποσχόμενη προσέγγιση παρόλα αυτά απαιτεί ακόμα ένα γονικό μοντέλο Transformer TTS ώστε να εκαπιδεύσει το non-autoregressive μοντέλο παραγωγής mel-spectrograms.

\section{Το πρόβλημα μετατροπής φωνής σε ανθρωπόμορφη οπτική ροή.}
Στην ενότητα αυτή θα θέλαμε ιδανικά να μελετήσουμε το πρόβλημα της οπτικοακουστικής σύνθεσης φωνής από κείμενο. Παρόλα αυτά στη βιβλιογραφία σπανίζουν οι δουλειές που μελετούν το συνολικό αυτό πρόβλημα. Ο λόγος που γίνεται αυτό είναι η πολύ αυξημένη πολυπλοκότητα του προβλήματος καθώς στην ουσία του λαμβάνουν μέρος οι ακόλουθοι μετασχηματισμοί 
\begin{itemize}
    \item μετατροπή κειμένου σε φασματογράφημα
    \item μετατροπή φασματογραφήματος σε κυματομορφή (ομιλία)
    \item μετατροπή ομιλίας σε ανθρωπόμορφο βίντεο
\end{itemize}
Ο σκοπός των δύο προηγούμενων ενοτήτων ήταν να περιγράψουμε τις μεθόδους που ακολουθούνται για να επιτευχθούν οι δύο πρώτοι μετασχηματισμοί. Σε αυτή την ενότητα θα επικεντρωθούμε κυρίως στο τρίτο πρόβλημα. 
Υπάρχουν διάφορες κατηγοριοποιήσεις ανάλογα με το πλαίσιο μελέτης, αλλά εμείς επιλέγουμε να χωρίσουμε τις προσεγγίσεις σε δύο επιμέρους. Αυτές που αποσκοπούν να απεικονίσουν ``αληθινά" (δεν παύουν να είναι συνθετικά) ανθρώπινα πρόσωπα και σε αυτές που εστιάζουν σε ανθρωπόμορφες φιγούρες όπως τα avatars.

\subsection{Προπαρασκευαστικές έννοοιες}
Στην ενότητα αυτή θα περιγράψουμε κάποιες βασικές έννοιες και κατευθύνσεις οι οποίες συναντώνται κατά τη μελέτη προβλημάτων σύνθεσης ανθρωπόμορφου προσώπου από φωνή. Η σημαντικότερη διαφορά μεταξύ των δύο είναι πως κατά τη σύνθεση ανθρωπόμορφων προσώπων όπως ένα avatar χρησιμοποιείται κάποιο μοντέλο προσώπου (face model) ενώ στη σύνθεση ρεαλιστικού προσώπου στοχεύουμε στην ανακασκευή του προσώπου ή τμημάτων αυτού. \par 
Πιο συγκεκριμένα κατά τη σύνθεση avatar χρησιμοποιούνται μοντέλα όπως τα AAΜ \cite{cootes1998AAM, cootes2001active, matthews2004active} και τα 3DMM \cite{blanz1999morphable, blanz2003face, egger20203d}. Τα ΑΑΜ παρότι αρκετά εύχρηστα δεν προτιμούνται πλέον στη βιβλιογραφία λόγω του ότι συνήθως παράγουν πιο ``φτωχούς" χώρους χαρακτηριστικών, έναντι των των λεγόμενω 3D Morphable Face Models (3DMM). 

\subsection{Οπτική σύνθεση ανθρωπόμορφης φιγούρας από φωνή}
\begin{figure}[!htb]
    \centering
    \includegraphics[scale=0.5]{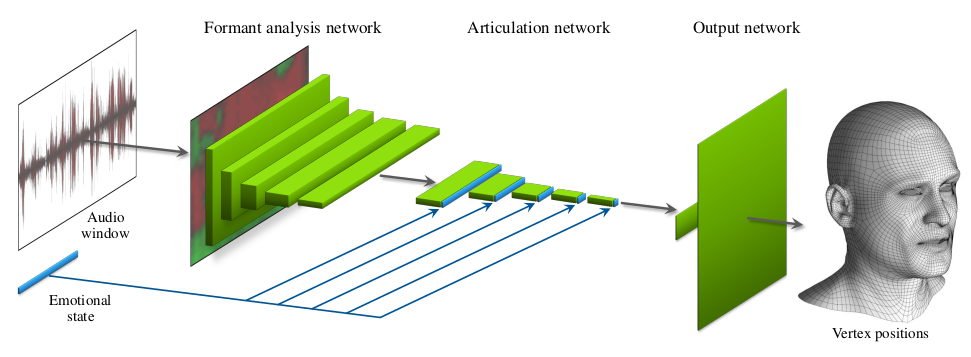}
    \caption{Karras acrhitecture \cite{karras_audio-driven_2017}}
    \label{fig:karras}
\end{figure}

\noindent
\textbf{Audio-Driven Facial Animation}. 
Η πρώτη σημανιτκή δουλειά που συνδύασε τις σύγχρονες μεθόδους deep learning και πέτυχε σημαντικά αποτελέσματα είναι από την Nvidia \cite{karras_audio-driven_2017}. Oι συγγραφείς αυτής της εργασίας ουσιαστικά εισήγαγαν μια μέθοδο που κάνει animate 3D ανθρωπόμορφα πρόσωπα. Συγκεκριμένα το δίκτυο τους (σχήμα \ref{fig:karras}) μαθαίνει μια απεικόνιση από την κυματομορφή εισόδου σε τρισδιάστατες vertex coordinates (συντεταγμένες κορυφών) ενός μοντέλου προσώπου. Η προσέγγιση τους εστιάζει στην περιοχή του προσώπου και όχι στα χείλη/στόμα όπως συνηθίζεται από αρκετές άλλες εργασίες. \par
Προκειμένου να διαφοροποιηθούν οι διάφορες εκφραστικές διαφορές που δεν μπορούν να εξαχθούν από το ηχητικό σήμα απευθείας χρησιμοποιείται και ένας επιπλέον latent (λανθάνων-κρυφός) κώδικας ο οποίος διαισθητικά μοντελοποιεί πιθανές συναισθηματικές διαφοροποιήσεις στην έκφραση. Κατά τη διάρκεια της σύνθεσης αυτός ο χώρος μπορεί να χρησμοποιηθεί προκειμένου να αποδωθούν διαφορετικές εκφράσεις στο παραγώμενο πρόσωπο.\par
Η αρχιτεκτονική προσέγγιση περιέχει ένα Formant Analysis Network το οποίο παράγει χρονομεταβλητές ακολουθίες από φωνητικά χαρακτηριστικά, τα οποία θα καθοδηγήσουν την άρθρωση (articulation). Στη συνέχεια ακολουθεί το Articulation δίκτυο το οποίο βρίσκει τη χρονική εξέλιξη των χαρακτηριστικών εκείνων που περιγράφουν τις κινήσεις του προσώπου. Τέλος ένα δίκτυο εξόδου παράγει τις τελικές μεταβλητές που ουσιαστικά ελέγχουν τις κινήσεις του προσώπου.\par 
Τα δεδομένα που χρησιμοποιήθηκαν έχουν ληφθεί ουσιαστικά βιντεοσκοπώντας ανθρώπους και φτιάχνοντας με λογισμικό τα ανθρωπόμορφα πρόσωπα. Για την επίτευξη ρεαλιστικών αποτελεσμάτων χρησιμοποιήθηκαν 3 διαφορετικές συναρτήσεις κόστους εκ των οποίων μία ( position) είναι υπεύθυνη για τη συνολική μετατόπιση κάθε vertex, μια (motion) για να παράγει ρεαλιστική κίνηση και μια που  λειτουργεί ως regularizer. Συγκεφαλαιώντας η προσέγγιση αυτή \cite{karras_audio-driven_2017} πετυχαίνει να παράξει ρεαλιστικά τρισδιάστατα πρόσωπα που βασίζονται αποκλειστικά σε ακουστική πληροφορία.\par
\begin{figure}[!htb]
    \centering
    \includegraphics[scale=0.5]{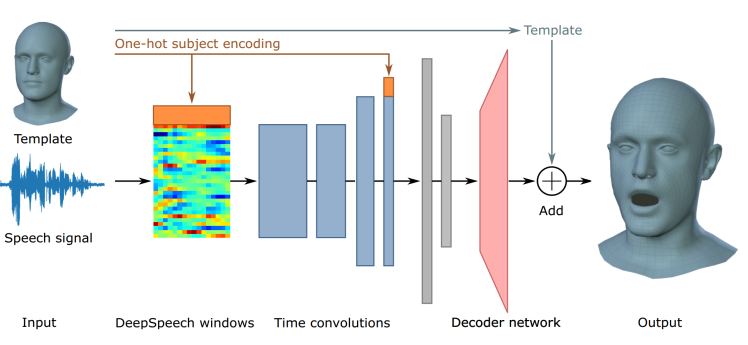}
    \caption{VOCA Architecture \cite{cudeiro_capture_2019}}
    \label{fig:voca}
\end{figure}

 
\vspace{5pt}
\noindent \textbf{Voice Operated Character Animation (VOCA)} Mια προσέγγιση στο ίδιο μήκος κύματος με την προηγούμενη \cite{karras_audio-driven_2017} παρουσιάστηκε στο CVPR 2019 υπό το όνομα VOCA \cite{cudeiro_capture_2019}. H συγκεκριμένη αρχιτεκτονική (σχήμα \ref{fig:voca}) δέχεται σαν είσοδο οποιοδήποτε σήμα φωνής και παράγει ρεαλιστικά 3D vertex animations. Παρέχει επίσης ένα controller (one-hot subject encoding) που μπορεί να αλλάζει το ύφος, την πόζα και άλλα χαρακτηριστικά για τον ομιλητή.\par
Σε αντίθεση με τον Karras \cite{karras_audio-driven_2017} που μαθαίνει ιδιοσυγκρασίες ενός ομιλητή, το VOCA είναι conditioned στα υποκείμενα (subjects) κατά τη διάρκεια της εκπαίδευσης και συνεπώς μπορεί να γενικεύει καλύτερα σε νέους ομιλητές. Η αρχιτεκτονική χρησιμοποιεί fetaures  που βγαίνουν από ένα pre-trained DeepSpeech \cite{hannun2014deep, amodei2016deep} μοντέλο. Στη συνέχεια τροφοδοτούνται στον encoder ο οποίος μαθαίνει ένα χώρο χαμηλής διάστασης και από αυτόν ένας decoder (αρχιτεκτονική με strided convolutions) είναι υπεύθυνος για την παραγωγή των τελικών θέσεων (ή για την ακρίβεια μετατοπίσεων) του 3D vertex. Με φιλοσοφία αρκετά παρόμοια με το \cite{karras_audio-driven_2017} χρησιμοποιεί επίσης πολλαπλά loss functions. Αξίζει ακόμα να αναφερθεί πως οι συγγραφείς του VOCA παρέχουν επίσης και τα scanned δεδομένα που χρησιμοποίησαν για να εκπαιδεύσουν το μοντέλο.\par 
\vspace{10pt}
\noindent \textbf{Taylor 2017} Στην εργασία αυτή \cite{taylor_deep_2017} παρουσιάζεται μια παρόμοια προσέγγιση με τις προηγούμενες δύο, μόνο που στην περίπτωση δεν μοντελοποιείται ολόκληρο το πρόσωπο του χαρακτήρα αλλά το κάτω μέρος του (μέσω ΑΑΜ μοντέλου). Επιπλέον παράγει μη εκφραστικά-ουδέτερα πρόσωπα για τον χαρακτήρα. Η αρχιτεκτονική που χρησιμοποιείται είναι ένα sliding window DNN (βαθύ νευρωνικό δίκτυο). Επίσης κατά τη διάρκεια της σύνθεσης χρειάζεται να χρησιμοποιηθεί ένα off-the-self (προεκπαιδευμένο) ASR σύστημα προκειμένου να μετατρέψει το σήμα εισόδου σε φωνήματα. Τα δεδομένα πάνω στα οποία εκπαιδεύτηκε το μοντέλο είναι ένα οπτικο-ακουστικό dataset με προτάσεις οι οποίες έχουνε λεχθεί με ουδέτερο τόνο.
\begin{figure}[!htb]
    \centering
    \includegraphics[scale=0.35]{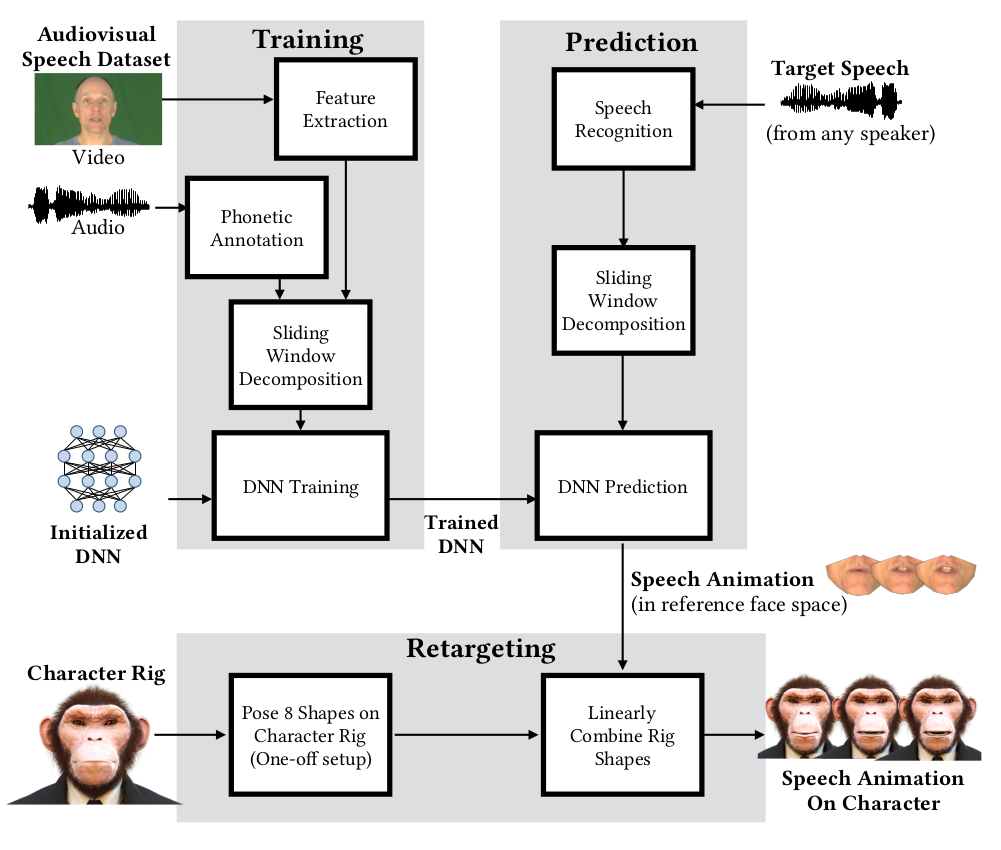}
    \caption{Taylor Architecture\cite{taylor_deep_2017}}
    \label{fig:taylor}
\end{figure}
\par 
\vspace{10pt}
\noindent \textbf{VisemeNet} Αυτή η προσέγγιση εισάγει το VisemeNet \cite{zhou2018visemenet}, ένα δίκτυο που αποτελείται από τρία στάδια, καθένα από τα οποία βασίζεται σε LSTMs \cite{hochreiter1997long}. Συγκεκριμένα περιέχει ένα landmark στάδιο, ένα phoneme-group στάδιο καθώς και ένα viseme στάδιο. Συνοπτικά η λειτουργία του δικτύου έχει ως εξής. Ο ήχος μοντελοποιείται με διανυσματικές αναπαραστάσεις διάστασης 1056, οι οποίες με τη σειρά τους τροφοδοτούνται στα στάδια landmark και phoneme τα οποία είναι υπεύθυνα για να προβλέψουν τη μετατόπιση (displacement) και το αντίστοιχο φώνημα. Τέλος τα διανύσματα εξόδου των δύο πρώτων σταδίων μαζί με την αρχική ηχητική αναπαράσταση δίνονται σαν είσοδος στο viseme στάδιο το οποίο προβλέπει τις παραμέτρους προσομοίωσης του ανθρωπόμορφου χαρακτήρα.
\par 
\vspace{10pt}
\noindent \textbf{Speech-driven 3D Facial Animation with Implicit Emotional Awareness} Αυτή η εργασία \cite{pham2017speech} προτάθηκε για παραγωγή τρισδιάστατων 3D ανθρωπομορφικών προσώπων σε πραγματικό χρόνο. Συγκεκριμένα, κατά τη διάρκεια της εκπαίδευσης εξάγονται τόσο οπτικά όσο και ακουστικά χαρακτηριστικά με τα οποία εκπαιδεύεται η LSTM-based αρχιτεκτονική (είσοδος ο ήχος, έξοδος τα οπτικά χαρακτηριστικά) που προβλέπει την περιστροφή της κεφαλής καθώς και την έκφραση του προσώπου ενός ομιλητή. Οι συγγραφείς αναφέρουν πως για την εκπαίδευση του μοντέλου χρειάζεται ένα μεγάλο οπτικοακουστικό σύνολο δεδομένων.

\subsection{Οπτική σύνθεση ανθρώπινου προσώπου από φωνή}
\noindent \textbf{Synthetizing Obama} H πρώτη εργασία \cite{suwajanakorn_synthesizing_2017} που πέτυχε εντυπωσιακά αποτελέσματα σε ότι αφορά τη σύνθεση πραγματικού βίντεο από φωνή. H προσέγγιση αφορά την παραγωγή ενός και μόνο προσώπου για το οποίο υπαχουν αρκετές διαθέσιμες βιντεοσκοπημένες ώρες (περίπου 17 ανφέρει η πρωτότυπη εργασία).
\begin{figure}[!htb]
    \centering
    \includegraphics[scale=0.4]{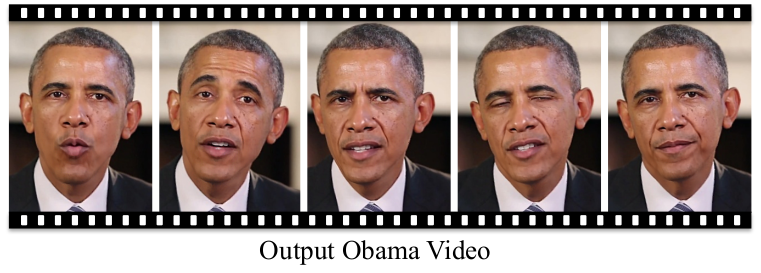}
    \caption{Synthetizing Obama paper \cite{suwajanakorn_synthesizing_2017} αποτελέσματα}
    \label{fig:obama}
\end{figure}
Ο αλγόριθμος περιέχει ένα χειροκίνητο βήμα κατά το οποίο επιλέγεται η περιοχή του στόματος και των δοντιών. Τα ακουστικά χαρακτηριστικά που χρησιμοποιούνται είναι MFCCs, τα οποία είναι μια ``γενική" περιγραφή για το σήμα. Οι συγγραφείς αναφέρουν ότι ιδανικά θα μπορούσε να χρησιμοποιηθεί απευθείας η κυματομορφή προκειμένου να εξαχθούν πιο χρήσιμα χαρακτηριστικά. Το RNN που χρησιμοποιείται προβλέπει τις μετατοπίσεις του στόματος καθώς και τις θέσεις των δοντιών. Τα αποτελέσματα τις μεθόδου οπτικοποιούνται στο Σχήμα \ref{fig:obama}. \par

\vspace{5pt}
\noindent \textbf{Realistic Speech-Driven Facial Animation with GANs} Σε αυτή την εργασία \cite{vougioukas_realistic_2019} χρησιμοποιούνται GANs για την παραγωγή ρεαλιστικών βίντεο ανθρώπων από φωνή.  
\begin{figure}[!htb]
    \centering
    \includegraphics[scale=0.45]{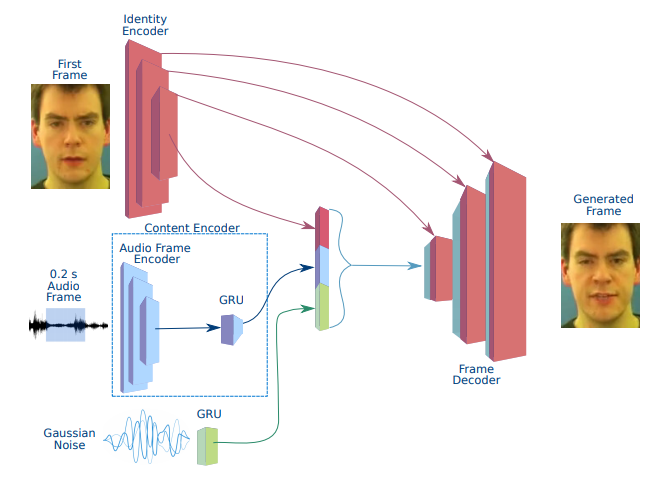}
    \caption{Facial Animation with GANs Generator \cite{vougioukas_realistic_2019}}
    \label{fig:vougioukas}
\end{figure}
Σε αντίθεση με προγενέστερες προσεγγίσεις που χρησιμοποιούν ενδιάμεσες visual αναπαραστάσεις και από εκεί μεταβαίνουν σε κάποια ανθρωπόμορφη αναπαράσταση (είτε avatar είτε πραγματικό πρόσωπο), η παρούσα εργασία παράγει συνθετικά ανθρώπινα πρόσωπα απευθείας από το ηχητικό σήμα.
Ένα σημαντικό διαφορετικό χαρακτηριστικό της μεθόδου έναντι της \cite{suwajanakorn_synthesizing_2017} είναι το γεγονός πως μοντελοποιεί ολόκληρο το πρόσωπο και όχι μόνο την περιοχή του στόματος.
Σε ότι αφορά την αρχιτεκτονική έχουμε ένα GAN με 3 discriminators καθένας εκ των οποίων αναλαμβάνει ένα διαφορετικό έργο (λεπτομέρειες, συγχρονισμός φωνής-εικόνας και ρεαλιστικές εκφράσεις). Σε ότι αφορά τον generator τώρα, έχουμε ένα κωδικοποιητή για την ταυτότητα του ομιλητή, έναν για το content καθώς και μια γεννήτρια παραγωγής τυχαίου θορύβου. Το παραγόμενο βίντεο είναι χαμηλότερης ποιότητας αλλά πολύ πιο ευέλικτο και μπορεί να χρησιμοποιηθεί για νέους ομιλητές.\par
\begin{figure}[!htb]
    \centering
    \includegraphics[scale=0.4]{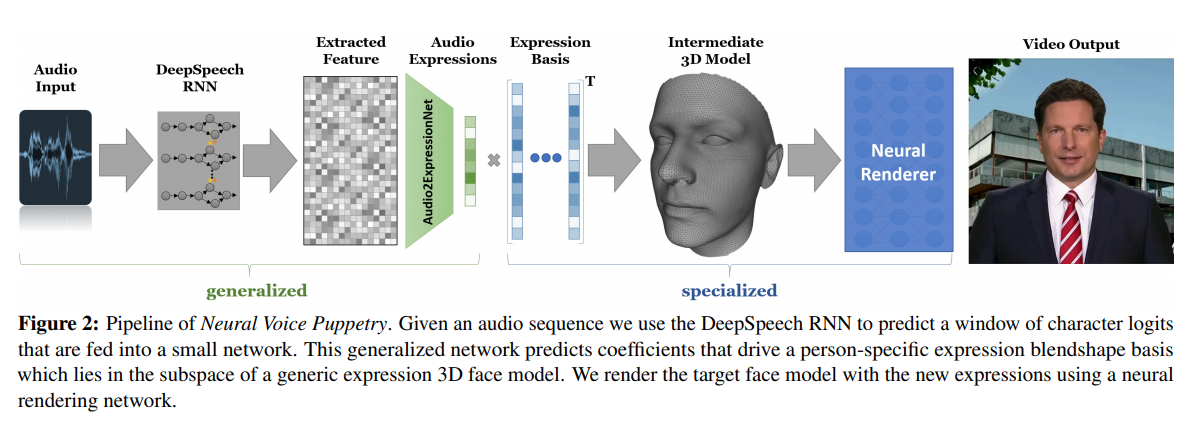}
    \caption{NVP Αρχιτεκτονική \cite{thies_neural_2019}}
    \label{fig:nvp}
\end{figure}

 
\vspace{5pt}
\noindent \textbf{Neural Voice Pupperty (NVP)} Πρόκειται για μια νέα προσέγγιση \cite{thies_neural_2019} για τη σύνθεση βίντεο προσώπου που βασίζεται στον ήχο. Λαμβάνοντας υπόψη ένα ηχητικό σήμα από ένα άτομο ή ψηφιακό βοηθό, ο αλγόριθμος δημιουργεί ένα φωτο-ρεαλιστικό βίντεο εξόδου ενός ατόμου-στόχου που είναι συγχρονισμένο με τον ήχο της εισόδου. Η ηχητική αναπαράσταση στην είσοδο, τροφοδοτείται σε ένα βαθύ νευρωνικό δίκτυο που μαθαίνει ένα latent χώρο για ένα 3D χωρικό μοντέλο. Μέσω της υποκείμενης τρισδιάστατης αναπαράστασης, το μοντέλο μαθαίνει εγγενώς τη χρονική σταθερότητα, ενώ αξιοποιούμε την υψηλή απόδοση του νευρωνικού δικτύου, για τη δημιουργία φωτο-ρεαλιστικών πλαισίων εξόδου. Η προσέγγιση αυτή γενικεύεται σε διαφορετικά άτομα, επιτρέποντάς μας να συνθέσουμε βίντεο ενός ανθρώπου (ή avatar) στόχου με  οποιαδήποτε φωνή ή ακόμα και συνθετικών φωνών που μπορούν να δημιουργηθούν χρησιμοποιώντας τυπικές προσεγγίσεις TTS. Η συνολική αρχιτεκτονική που περιγράψαμε φαίνεται και στο Σχήμα \ref{fig:nvp}.

\subsection{MakeItTalk}
H δουλειά αυτή είναι ίσως η πιο σημαντική των τελευταίων ετών πάνω σε οπτική σύνθεση οδηγούμενη από φωνή. Πρόκειται για εργασία της Adobe με τίτλο \textit{MakeItTalk} \cite{zhou2020makelttalk}. Αυτό που πετυχαίνει είναι ότι δοθέντος μιας εικόνας-πορτρέτου ενός ανθρώπου ή cartoon και ενός ηχητικού σήματος παράγει ένα βίντεο που προσομειώνει το χαρακτήρα να μιλάει. 

\begin{figure}[!htb]
    \centering
    \includegraphics[scale=0.15]{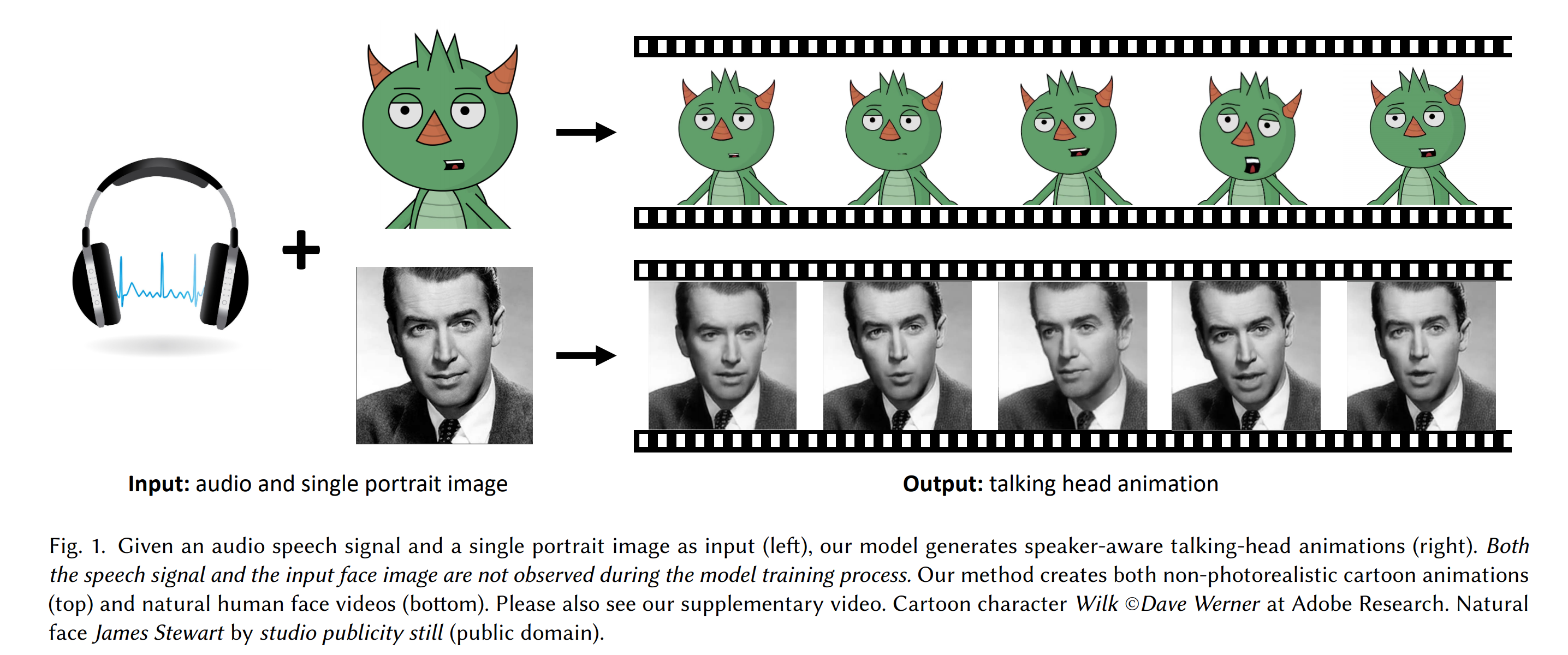}
    \caption{MakeItTalk εποπτική λειτουργία \cite{zhou2020makelttalk}}
    \label{fig:mkittalk_1}
\end{figure}

Στα πολύ θετικά αυτής της προσέγγισης είναι ότι δεν χρησιμοποιεί κάποιο μοντέλο προσώπου (face model) ώστε να μιμηθεί την κίνηση  ενός πραγματικού προσώπου. Ανταυτού χρησιμοποιεί κάποιες πολύ μεθοδευμένες προσεγγίσεις οι οποίες δίνουν τρισδιάστατα facial landmarks. Συγκεκριμένα για ανθρώπινα πρόσωπα χρησιμοποιείται ο state-of-the-art αλγόριθμος \cite{bulat2017far}, ενώ για cartoon ή σκίτσα έχει χρησιμοποιηθεί ο αλγόριθμος Face-of-Art \cite{yaniv2019face}, ο οποίος είναι ειδικά σχεδιασμένος για να αναγνωρίζει landmarks σε πίνακες ζωγραφικής.

Η άλλη εξίσου σημαντική υπόθεση που κάνει η συγκεκριμένη προσέγγιση είναι ότι χρησιμοποιεί ένα voice conversion αλγόριθμο, π.χ \cite{qian2019autovc}, προκειμένου να αποσυμπλέξει την πληροφορία του ηχητικού σήματος αναφορικά με τον ομιλητή (speaker embedding) και το περιεχόμενο (content embedding). Με τον τρόπο αυτό προβλέπουν τις μετατοπίσεις των landmarks και τις προσαρμόζουν κάθε φορά στο στυλ του εκάστοτε ομιλητή.

Η τελική σύνθεση του προσώπου γίνεται με βάση τις προβλεπόμενες μετατοπίσεις από το δίκτυο και είτε ένα Image2Image translation δίκτυο για τις εικόνες ανθρώπων, είτε με βάση τον αλγόριθμο τριγωνοποίησης Delaunay για τα cartoon.

\begin{figure}[!htb]
    \centering
    \includegraphics[scale=0.15]{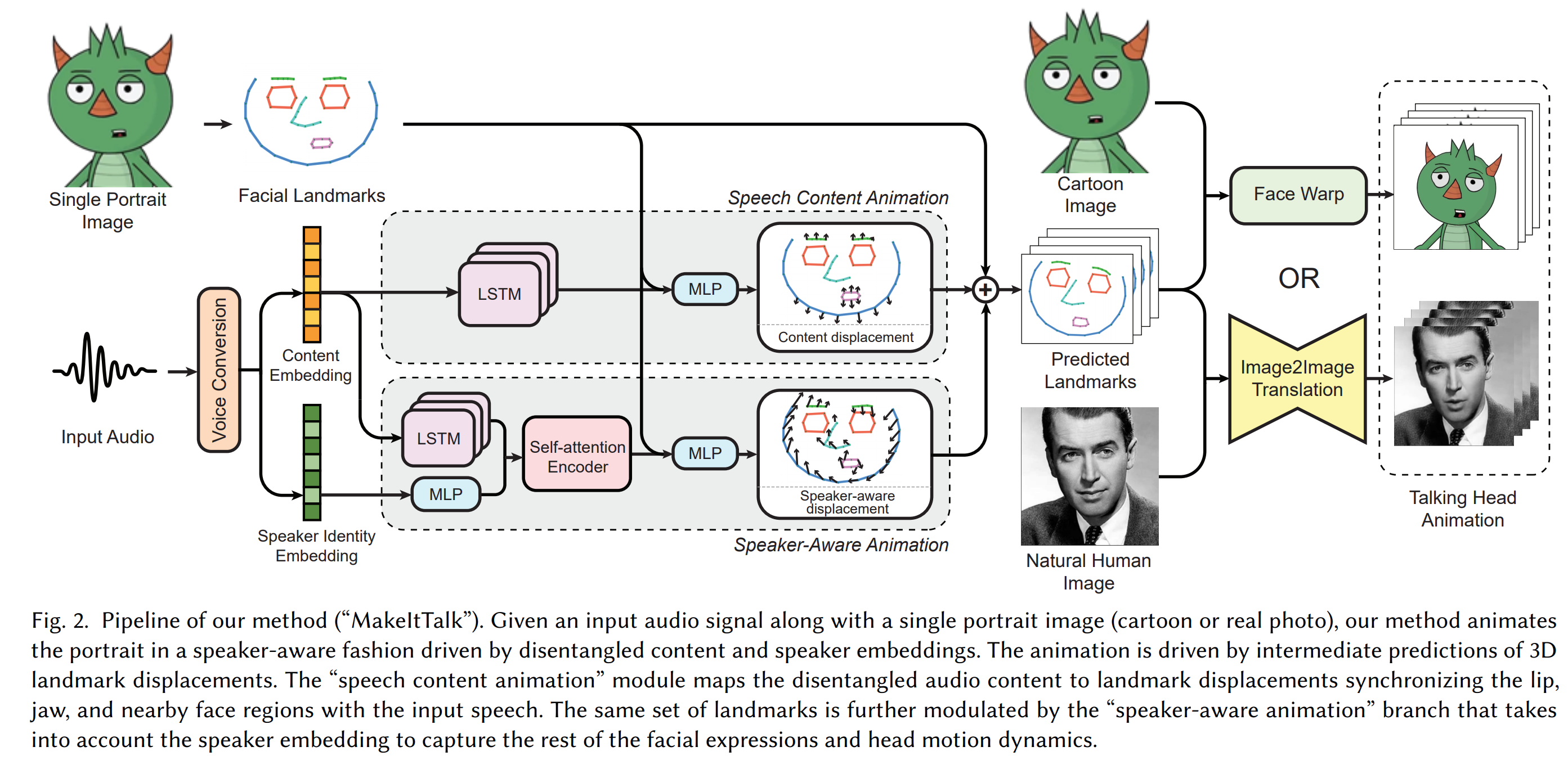}
    \caption{Αρχιτεκτονική Μοντέλου \cite{zhou2020makelttalk}}
    \label{fig:model_architecture}
\end{figure}

Aπό αρχιτεκτονικής σκοπιάς, χρησιμοποιεί LSTM και κάποιους μηχανισμούς self-attention ως εκαπιδευόμενα τμήματα του δικτύου, καθώς επίσης και MLPs που χρησιμοποιούνται κυρίως ως fusion ή dimensionality reduction υποδίκτυα.

Στα αρνητικά της προσέγγισης είναι ο υψηλός χρόνος που απαιτείται για να συντεθεί το τελικό avatar-cartoon. Επίσης τα αποτελέσματα της σύνθεσης σε πραγαμτικά πρόσωπα είναι συγκρίσιμα μεν αλλά υποδεέστερα έναντι κάποιων που βασίζονται σε blendshapes.

\subsection{Άλλες Προσεγγίσεις}
Στην ενότητα αυτή περιγράφουμε εργασίες οι οποίες δεν εμπίπτουν άμεσα στην παραπάνω κατηγοριοποίηση καθώς επίσης παραθέτουμε και μια σύντομη συζήτηση πάνω στις παρούσες προσεγγίσεις και πιθανούς περιορισμούς τους. Τέλος κάνουμε μια σύντομη ανασκόπηση των ευρέως διαδεδομένων βάσεων δεδομένων που μπορούν να χρησιμοποιούν για την ανάπτυξη τέτοιων συστημάτων.
\newpage
\noindent \textbf{Λοιπές Εργασίες}
Η πιο σημαντική είναι η πιο πρόσφατη εργασία οπτικοακουστικής σύνθεσης στην Ελληνική \cite{filntisis2017video}. Η εργασία αυτή βρίσκεται στην τομή των σύγχρονων και των παλαιότερων προσεγγίσεων. Συγκεκριμένα προτείνονται δύο αρχιτεκτονικές βαθιού νευρωνικού δικτύου (DNN) για σύνθεση ρεαλιστικής εικόνας Expressive Audio-Visual Text-ToSpeech (EAVTTS), οι οποίες αξιολογούνται συγκρίνοντάς τις απευθείας και με το παραδοσιακό κρυφό μοντέλο Markov (HMM) EAVTTS, καθώς και μια συνδυαστική προσέγγιση EAVTTS unit selection, τόσο στον ρεαλισμό όσο και στην εκφραστικότητα των παραγόμενων talking heads.
\par 
Μια εργασία η οποία παρουσίασε εντυπωσιακά αποτελέσματα είναι το λεγόμενο First Order Motion \cite{siarohin_first_2019}. Στη δουλειά αυτή δοθέντος ενός βίντεο-καθοδηγητή (drive video) και μιας εικόνας-στόχου (target) παράγεται ένα βίντεο το οποίο μιμείται το αρχικό αλλά χρησιμοποιώντας την εικόνα στόχο. Η σημασία αυτής της εργασίας αναδεικνύεται αφενός λόγω της πληθώρας των εφαρμογών της από ανθρώπινα πρόσωπα μέχρι stickers, αφετέρου δε χρειάζεται κάποια ισχυρά priors όπως συμβαίνει συνήθως με landmark μοντέλα. Είναι δηλαδή άμεσα εφαρμόσιμη σε οποιαδήποτε φιγούρα. Συγγενής εργασία αποτελέι το MakeItTalk \cite{zhou2020makelttalk}

\vspace{5pt}
\noindent \textbf{Περί Μοντέλων Προσώπου}
Οι μέχρι στιγμής καλύτερες μέθοδοι που έχουμε διαθέσιμες, π.χ VOCA \cite{cudeiro_capture_2019}, NVP \cite{thies_neural_2019}, χρησιμοποιούν κάποια μοντέλα προσώπου. Αυτά έχουν κατασκευαστεί από διαθέσιμα 4Dscans που ουσιαστικά μοντελοποιούν πολλά σημεία του προσώπου σχετικά με τον τρόπο που κινούνται κατά την εκφορά συγκεκριμένων φωνημάτων, λέξεων και προτάσεων. Πολλές λοιπόν μέθοδοι βασίζονται πάνω σε τέτοιες αναπαραστάσεις του προσώπου και συγκεκριμένα τα λεγόμενα 3DMM \cite{blanz1999morphable, blanz2003face} που ουσιαστικά υποδεικνύουν την κίνηση ενός προσώπου με βάση μια πληθώρα παραμέτρων που καλούνται blendshapes. Οι τιμές αυτές ελέγχουν ουσιαστικά μια κίνηση στο πρόσωπο και ο γραμμικός τους συνδυασμός αναπαριστά το συνολικό αποτέλεσμα.
\par
Εαν τώρα θελήσουμε να εκπαιδεύσουμε κάποιο σύστημα οπτικοακουστικής σύνθεσης σε δικά μας δεδομένα τότε μια λύση είναι να βρούμε τις ground truth blenshape παραμέτρους. Η διαδικασία αυτή απαιτεί πολύ κοστοβόρο εξοπλισμό και προϋποθέτει ουσιαστικά να συλλέγουμε δεδομένα για κάθε εφαρμογή που επιθυμούμε. Κάτι τέτοιο φυσικά είναι αδύνατο και συνεπώς έχουν προταθεί λύσεις οι οποίες ουσιαστικά προσπαθούν να προσαρμόσουν ένα ήδη υπάρχον συστημα μοντέλων προσώπου στα δεδομένα μας. Υπάρχουν δύο προσεγγίσεις. Η πρώτη είναι optimization-based \cite{thies2016face2face} ενώ η δεύτερη βασίζεται σε unsupervised encoder-decoder αρχιτεκτονικές \cite{tewari2017mofa}. Για την εκπαίδευση λοιπόν του εκάστοτε μοντέλου απαιτείται πρώτα να εξάγουμε τις πραγματικές τιμές των blendshapes τις οποίες στη συνέχεια θα κληθούμε να δώσουμε σαν επιθυμητή έξοδο στο μοντέλο μας.

\vspace{5pt}
\noindent \textbf{Οπτικοακουστικές Βάσεις Δεδομένων}
Δεν θα πρέπει να αμελούμε το γεγονός πως όλες οι προσεγγίσεις που βασίζονται σε βαθιά νευρωνικά δίκτυα απαιτούν μεγάλο όγκο δεδομένων προκειμένου να λειτουργήσουν. Για το λόγο αυτό κρίνουμε σκόπιμο να αναφέρουμε επιγραμματικά μερικές από τις μεγαλύτερες βάσεις δεδομένων. Μερικά από αυτά τα δεδομένα είναι το VoxCeleb \cite{Nagrani17} και το VoxCeleb2 \cite{Chung18b} (το 2 είναι υπερσύνολο του πρώτου) που περιέχουν μικρής διάρκειας βίντεο από συνεντεύξεις διαθέσιμες στο διαδίκτυο. Στο ίδιο μήκος κύματος είναι και το AVSpeech \cite{ephrat2018looking} τo οποίο περιέχει βίντεο από το YouTube.

\section{Σύνοψη και Συμπεράσματα}

Ο σκοπός της παρούσας βιβλιγογραφικής ανασκόπησης ήταν να περιγράψει το πρόβλημα της οπτικοακουστικής σύνθεσης. Συγκεκριμένα προσπαθήσαμε να γίνει εξαρχής φανερό ότι το πρόβλημα αυτό αποτελεί αρκετά σύνθετο και θα πρέπει να χωριστεί και να μελετηθεί σε επιμέρους προβλήματα. Συγκεκριμένα, οι δύο βασικοί άξονες-προβλήματα είναι η σύνθεση φωνής (TTS) και η δημιουργία ανθρωπόμορφης οπτικής ροής οδηγούμενης από φωνή.

Ως προς τη σύνθεση φωνής συνήθως χρειάζεται να εκπαιδεύσουμε δύο επιμέρους δίκτυα. Το πρώτο θα απεικονίζει τη λεκτική περιγραφή-κείμενο σε κάποια ``ενδιάμεση'' αναπαράσταση, π.χ. φασματογραφήματα. Το δεύτερο, θα δέχεται σαν είσοδο αυτή την ενδιάμεση αναπαράσταση και θα την απεικονίζει στην κυματομορφή της φωνής. Αυτά τα δίκτυα που παράγουν την κυματομορφή καλούνται vocoders. Εξαιτίας της πολυπλοκότητας των vocoder δικτύων μεγάλος όγκος ερευνητικής δουλειάς είναι αφιερωμένος αποκλειστικά στη μελέτη και την ανάπτυξη τέτοιου έιδους δικτύων.

Το δεύτερο πρόβλημα που παρουσιάσαμε είναι η δημιουργία ανθρωπόμορφης οπτικής ροής οδηγούμενη από φωνή.
Προσεγγίσεις περιλαμβάνουν αλγόριθμους που συνθέτουν ανθρώπινα πρόσωπα, πλέγματα (meshes) ανθρώπινων προσώπων καθώς επίσης avatars και καρικατούρες. Οι περισσότερες προσεγγίσεις χρησιμοποιούν κάποιο μοντέλο προσώπου (face model). Τα μοντέλα αυτά ουσιαστικά είναι φτιαγμένα από human artists και χρησιμοποιούν έναν αριθμό παραμέτρων προκειμένου να ελέγχουν τις κινήσεις του προσώπου. Η σύνθεση του ανθρωπόμορφου προσώπου γίνεται συνήθως μέσω των παραμέτρων αυτών και τα μοντέλα που παρουσιάσαμε ουσιαστικά καλούνται να προβλέψουν τις παραμέτρους αυτές δοθέντος ενός ηχητικού σήματος. Ένας σημαντικός παράγοντας που θα πρέπει να ληφθεί υπόψιν είναι ο τρόπος που θα ανακτήσουμε τις παραμέτρους για τα εκάστοτε μοντέλα προσώπου, προκειμένου να εκαπιδέυσουμε το δίκτυο που θα κληθεί να τις προβλέψει. Άλλες προσεγγίσεις στη βιβλιογραφία βασίζονται σε πιο απλές μοντελοποιήσεις του ανθρωπίνου προσώπου όπως τα landmarks και τα AAM. Επίσης αρκετές προσεγγίσεις στη βιβλιογραφία χρησιμοποιούν προεκπαιδευμένα μοντέλα προκειμένου να εξάγουν χαρακτηριστικά από το ηχητικό σήμα.

Eν κατακλείδι, στο σχεδιασμό και την υλοποίηση ενός τέτοιου προβλήματος δεν υπάρχει μία αποκλειστικά προσέγγιση που να οδηγεί στο επιθυμητό αποτέλεσμα. Για το λόγο αυτό θα πρέπει να μελετάμε τόσο τα δεδομένα όσο και τις ``προδιαγραφές'' του συστήματος που θέλουμε να κατσκευάσουμε και στη συνέχεια να μεταβούμε στην υλοποίηση/τροποποίηση κάποιας αρχιτεκτονικής δικτύου.

\bibliographystyle{apalike}
\bibliography{references}

\end{document}